\documentclass[12pt]{article}

\ifx\pdfoutput\undefined
\usepackage[dvips,bookmarks]{hyperref}
\else
\usepackage{hyperref}
\fi
\hypersetup{colorlinks=false,bookmarksopen,bookmarksnumbered,citecolor=blue,
   pdfstartview=FitH}

\usepackage[pdftex]{graphicx}	
\usepackage{latexsym}

\usepackage{amssymb,amsfonts,amsmath}
\usepackage{graphicx} 
\usepackage{indentfirst}

 \usepackage{bbm}

\parskip=\medskipamount

\newcommand {\cC}{{\cal C}}
\newcommand {\cD}{{\cal D}}

\newcommand {\cK}{{\cal K}}
\newcommand {\cL}{{\cal L}}
\newcommand {\cM}{{\cal M}}
\newcommand {\cN}{{\cal N}}

\newcommand {\cR}{{\cal R}}
\newcommand {\cS}{{\cal S}}
\newcommand {\cT}{{\cal T}}

\newcommand{\bD}{{\bf D}}

\def\a{\alpha}

\def\b{\beta}

\def\d{\delta}

\def\g{\gamma}

\def\k{\kappa}
\def\l{\lambda}
\def\m{\mu}

\def\o{\omega}

\def\q{\theta}
\def\r{\rho}
\def\s{\sigma}
\def\t{\tau}

\def\x{\xi}

\def\O{\Omega}

\def\S{\Sigma}

\def\X{\Xi}
\def\tr{{\rm tr}}
\def\rd{{\rm d}}
\def\ri{{\rm i}}

\newcommand{\ve}{\varepsilon}                            

\newcommand{\pa}{\partial}                           
\newcommand{\hf}{\frac12}

\newcommand{\vf}{\varphi}

\newcommand{\be}{\begin{equation}}
\newcommand{\ee}{\end{equation}}
\newcommand{\bea}{\begin{eqnarray}}
\newcommand{\eea}{\end{eqnarray}}
\newcommand{\non}{\nonumber}
\newcommand{\bm}[1]{\mbox{\boldmath$#1$}}

\def\double #1{#1{\hbox{\kern-2pt $#1$}}}

\newif\ifdtup

\def\de{{\nabla}}                                         

\newcommand{\bsubeq}{\begin{subequations}}
\newcommand{\esubeq}{\end{subequations}}

\numberwithin{equation}{section}

\begin{document}
\begin{titlepage}
\begin{flushright}
December, 2012\\
\end{flushright}
\vspace{5mm}

\begin{center}
{\Large \bf 
Conformal supergravities as Chern-Simons theories revisited
}\\ 
\end{center}

\begin{center}

{\bf
Sergei M. Kuzenko
and
Gabriele Tartaglino-Mazzucchelli\footnote{gabriele.tartaglino-mazzucchelli@uwa.edu.au}
} \\
\vspace{5mm}

\footnotesize{
{\it School of Physics M013, The University of Western Australia\\
35 Stirling Highway, Crawley W.A. 6009, Australia}}  
\vspace{2mm}

\end{center}

\begin{abstract}
\baselineskip=14pt
We propose a superfield method to construct off-shell actions for $\cN$-extended conformal supergravity
theories in three space-time dimensions. It makes use of the superform technique to engineer 
supersymmetric invariants. 
The method is specifically applied to the case 
of $\cN=1$ conformal supergravity and provides a new realization for the actions of 
conformal and topologically massive supergravities.
\end{abstract}

\vfill

\vfill
\end{titlepage}

\newpage
\renewcommand{\thefootnote}{\arabic{footnote}}
\setcounter{footnote}{0}

%

\allowdisplaybreaks


\section{Introduction}
\setcounter{equation}{0}

Inspired by the construction of three-dimensional 
$\cN=1$ topologically massive supergravity \cite{DK,Deser},
 conformal supergravities in three dimensions were formulated as 
supersymmetric Chern-Simons theories for  $\cN=1$ \cite{vN}, $\cN=2$ \cite{RvN},
and finally for arbitrary $\cN$ \cite{LR89}.
The approaches pursued in \cite{vN,RvN,LR89} are purely component. 
They are on-shell for $\cN>2$, 
and do not allow for a conventional superspace reformulation for $\cN>1$. The action for $\cN=1$ conformal supergravity can readily be constructed in terms of the 
superfield connection as a superspace integral \cite{GGRS,ZP88,ZP89}
(although the results in  \cite{GGRS,ZP88,ZP89} are incomplete, see section 3).
However, such a construction becomes impossible starting from $\cN=2$.
This is because (i)  the spinor and vector sectors of the superfield connection 
have  positive dimension equal to $1/2$ and 1
respectively; 
and (ii) the dimension of the full superspace measure is  $(\cN- 3)$. 
As a result, it is not possible to construct 
contributions to the action 
that are cubic in the superfield connection for $\cN\geq 2$. 

Nevertheless, it turns out that $\cN$-extended conformal supergravity 
can be realized in terms of the off-shell 
Weyl supermultiplet \cite{HIPT} and the associated curved superspace geometry originally sketched 
in \cite{HIPT} and later fully developed in \cite{KLT-M11}. 
Such a realization is a generalization of the superform formulation for the linear multiplet in 
four-dimensional $\cN=2$ conformal supergravity given in \cite{BKN}.\footnote{At the heart 
of the formulation \cite{BKN} lies the superform approach to 
the construction of supersymmetric invariants \cite{Hasler,Ectoplasm,GGKS}, also know as 
 the ectoplasm formalism \cite{Ectoplasm,GGKS}.} 
In the present paper we will first describe our method by 
explicitly constructing a new action principle for $\cN=1$ conformal supergravity in three dimensions. 
After doing so, an outline will be given as to how this method should be used
in the case of  extended conformal supergravities. 

Three-dimensional $\cN=1$ supergravity is an old topic that goes back to 1977. 
Without pretending to give a complete literature review, 
here we only list several  works
\cite{HT,BG,Uematsu} which initiated this research topic. 

This paper is organized as follows. In section 2 we review the superspace 
geometry of $\cN=1$ conformal supergravity. 
In section 3 we describe standard actions  for conformal supergravity and topologically massive supergravity realized as superspace integrals. 
Here we also present the $\cN=1$ supersymmetric Cotton tensor.  
Section 4 describes our main construction.  	
A component analysis of our results is given in section 5. 
In section 6 we give a sketch of our method as applied to 
extended conformal supergravity. The main body of the paper is accompanied by  two technical appendices concerning a locally supersymmetric action and
exact three-forms in superspace.


\section{$\cN=1$ conformal supergravity in superspace}
\label{section2}
\setcounter{equation}{0}

In this section we review the geometry of $\cN=1$ conformal supergravity in superspace
following the notation and conventions of \cite{KLT-M11}.
Let us consider a curved 
superspace, which is locally parametrized by real   bosonic ($x^m$) 
and real fermionic ($\q^{\mu}$) coordinates
\bea
z^{M}=(x^m,\q^{\mu})~,
 \qquad m=0,1,2
~,~~~
\mu=1,2
~.
\eea
The superspace structure group is the double covering of the Lorentz group, 
${\rm SL}(2,{\mathbb R})$,  
and we denote by $\cM_{ab}=-\cM_{ba}$ the Lorentz generators.
The covariant derivatives have the form:
\bea
\cD_{A}&=& (\cD_a, \cD_\a )= E_{A}+\O_A~.
\eea
Here the vector fields $E_A=E_A{}^M 
\pa/\pa z^M$ define the inverse vielbein,
and 
\bea
\O_A=\hf\O_{A}{}^{bc}\cM_{bc}=-\O_{A}{}^b\cM_b=\hf\O_{A}{}^{\b\g}\cM_{\b\g}~,
~~~~
\cM_{ab}=-\cM_{ba}~,~~\cM_{\a\b}=\cM_{\b\a}
\eea
is the Lorentz connection.
The Lorentz generators with two vector indices ($\cM_{ab}$), with one vector index ($\cM_a$)
and with two spinor indices ($\cM_{\a\b}$) are related to each other by the rules:
$\cM_a=\hf \ve_{abc}\cM^{bc}$ and $\cM_{\a\b}=(\g^a)_{\a\b}\cM_a$.

The supergravity gauge group is generated by local transformations of the form
\bea
\d_\cK\cD_A=[\cK,\cD_A]~,~~~~~~
\cK=\x^C E_C+\hf K^{cd} \cM_{cd}
~,
\label{SUGRA-gauge-group1}
\eea
with all the gauge parameters obeying natural reality conditions but otherwise arbitrary.
Given a tensor superfield $T$, it transforms as follows:
\bea
\d_\cK T=\cK T ~.
\label{SUGRA-gauge-group2}
\eea

The covariant derivatives satisfy the (anti)commutation relations
\bea
{[}\cD_{{A}},\cD_{{B}}\}&=&T_{{A}{B}}{}^{{C}}\cD_{{C}}
+\hf R_{{A}{B}}{}^{cd}\cM_{cd}
~,
\label{algebra-4-2-N}
\eea
with  $T_{AB}{}^C$ the torsion and  $R_{AB}{}^{cd}$ the Lorentz curvature. 
Similar to the connection, 
the Lorentz curvature can be realized in three different forms as  tensors carrying
 two vector indices ($R_{AB}{}^{cd}$), 
one vector index ($R_{AB}{}^{c}$)
and two spinor indices ($R_{AB}{}^{\g\d}$). These  are related to each other by the rules:
$R_{AB}{}^{c}=\hf \ve^{cde}R_{AB}{}_{de}$ and $R_{AB}{}^{\g\d}=(\g_c)^{\g\d}R_{AB}{}^{c}$.

To describe conformal supergravity, the covariant derivatives have to obey certain constraints
such that the complete algebra of covariant derivatives, 
compatible with all the Bianchi identities, is
\bsubeq
\bea
\{\cD_\a,\cD_\b\}&=&
2\ri\cD_{\a\b}
-4\ri \cS\cM_{\a\b}
~,~~~~~~~~~
\label{N=1alg-1}
\\
{[}\cD_{a},\cD_\b{]}
&=&
\cS(\g_a)_\b{}^{\g}\cD_{\g}
-(\g_a)_\b{}^{\g}C_{\g\d\r}\cM^{\d\r}
-\frac{2}{3}\big(
(\cD_{\b}\cS)\d_a^c
-2\ve_{ab}{}^{c}(\g^b)_{\b\g}(\cD^{\g}\cS)\big)\cM_c
~,~~~~~~~~~
\label{N=1alg-3/2-2}
\\
{[}\cD_{a},\cD_b{]}
&=&
\ve_{abc}\Big{\{}
\Big{[}\frac{\ri}{2}(\g^c)_{\a\b}C^{\a\b\g}
-\frac{2\ri}{3}(\g^c)^{\b\g}(\cD_{\b}\cS)\Big{]}\cD_\g
\non\\
&&
~~~~~~
-\Big{[}
-\frac{\ri}{2}(\g^c)^{\a\b}(\g^d)^{\g\d}\cD_{(\a}C_{\b\g\d)}
-\Big(
\frac{2\ri}{3}(\cD^2\cS)
+4\cS^2\Big)\eta^{cd}
\Big{]}\cM_d
\Big{\}}
~.
~~~~~~~~~~~~
\label{N=1alg-2}
\eea
\esubeq
Here the scalar $\cS$ is real and the spinor $C_{\a\b\g}=C_{(\a\b\g)}$ is real and 
completely symmetric and $\cD^2 := \cD^\a \cD_\a$.
The dimension-2 Bianchi identities imply that 
\bea
\cD_{\a}C_{\b\g\d}&=&
\cD_{(\a}C_{\b\g\d)}
-\ri\ve_{\a(\b}\cD_{\g\d)}\cS
~,
\eea
and hence 
\bea
\cD^{\g}C_{\a\b\g} = -\frac{4\ri}{3}  \cD_{\a\b} \cS~.
\label{2.9}
\eea 

The algebra of covariant derivatives is invariant under 
arbitrary super-Weyl transformations \cite{ZP88,ZP89,LR-brane} of the form 
\bsubeq
\bea
\d_\s\cD_\a&=&
\hf \s\cD_\a + (\cD^{\b}\s)\cM_{\a\b}
~,
\\
\d_\s\cD_a&=&
\s\cD_a
+{\ri\over 2}(\g_a)^{\g\d}(\cD_{\g} \s)\cD_{\d}
+\ve_{abc}(\cD^b\s)\cM^{c}
~,
\eea
\esubeq
with the parameter $\s$ being a real unconstrained superfield, provided 
the torsion superfields transform as
\bsubeq
\bea
\d_\s\cS&=&\s\cS-\frac{\ri}{4}  \cD^2\s~,
\label{sWS}
\\
\d_\s C_{\a\b\g}&=&\frac{3}{2}\s C_{\a\b\g}-\hf  \cD_{(\a\b}\cD_{\g)}\s
~.
\label{sWC}
\eea
\esubeq
The super-Weyl invariance is compulsory  for the above geometry to describe the multiplet of 
$\cN=1$ conformal supergravity. This local symmetry will play a crucial role in the present paper.

We introduce the vielbein 
and connection one-forms defined by
\bsubeq
\bea
E^A&:=&\rd z^ME_{M}{}^A
~,
\\
\O_C{}^D&:=&\rd z^M\O_{M}{}_C{}^D=E^A\O_A{}_C{}^D
~, 
\eea
\esubeq
where the supermatrix $E_M{}^A$ is the inverse of $E_A{}^M$,
\bea
E_M{}^AE_A{}^N=\d_M^N~,~~~E_A{}^ME_M{}^B=\d_A^B
~.
\eea
The connection one-form is associated with the Lorentz group, 
\bea
\O_A{}^B=
\left(
\begin{array}{c|c}
\O_a{}^b &0\\
\hline
0& \hf\O_\a{}^\b
\end{array}
\right)
=\left(
\begin{array}{c|c}
-\ve_{da}{}^b\O^d &0\\
\hline
0& \hf(\g_d)_\a{}^\b\O^d
\end{array}
\right)
~.
\eea

The superspace geometry of $\cN=1$ conformal supergravity can be recast 
in terms of superforms which will be a crucial ingredient of our construction.
The torsion and the curvature two-forms are 
\bsubeq
\bea
T^C&:=&\hf E^B\wedge E^AT_{AB}{}^C=-\rd E^C+E^B\wedge\O_B{}^C
~,
\\
R_C{}^D&:=&\hf E^B\wedge E^AR_{AB}{}_C{}^D=\rd\O_C{}^D-\O_C{}^E\wedge\O_E{}^D
~.
\eea
\esubeq
Here we have explicitly indicated the operation of wedge product ($\wedge$) of superforms. 
However, in the remainder of the paper it will be assumed and not be given explicitly.  
Given a $p$-form 
$F_p:=\frac{1}{p!}\rd z^{M_p}\cdots \rd z^{M_1}F_{M_1\cdots M_p}$ 
its exterior derivative can be written in two different forms:
\bsubeq
\bea
\rd F_p&=&\frac{1}{p!}\rd z^{M_p} \cdots \rd z^{M_1}  \rd z^N
\pa_NF_{M_1\cdots M_p}
\\
&=&
\frac{1}{p!} E^{A_p}\cdots E^{A_1}E^B\Big\{\cD_{B}F_{A_1\cdots A_p}
-\frac{p}{2}T_{BA_1}{}^CF_{CA_2\cdots A_p}\Big\}
~.
\eea
\esubeq

For the subsequent analysis we will need the super-Weyl transformations of the 
vielbein and the connection one-forms. They are
\bsubeq
\bea
\d_\s E^\a&=&
-\hf\s E^\a
-\dfrac{\ri}{2}(\g_b)^{\a\b}(\cD_\b\s)E^b 
~,
\label{sWE-1/2}
\\
\d_\s E^a&=&
-\s E^a
~,
\label{sWE-1}
\eea
\esubeq
and
\bea
\d_\s\O{}^{c}&=&
E^\a\Big\{-(\g^c)_{\a\b}(\cD^{\b}\s)\Big\}
+E^{a}\Big\{-\ve_{ab}{}^c(\cD^{b}\s)
\Big\}
~.
\label{2.18}
\eea

To construct a locally supersymmetric action principle, 
we need a real scalar Lagrangian $\cL$, of mass dimension $+2$,  with the  super-Weyl 
transformation law \cite{KLT-M11}
\bea
\d_\s\cL=2\s\cL~.
\label{N=1sWL}
\eea
The action is
\bea
S&=& \ri \int\rd^3x\rd^2\q E \,\cL
~,~~~
\qquad E^{-1}= {\rm Ber}(E_A{}^M)
~.
\label{N=1Ac}
\eea
The action is super-Weyl invariant, since the super-Weyl transformation of $E$ 
proves to be  $\d_\s E=-2\s E$.
Instead of defining the action principle using the superspace integration, 
 it suffices to construct a three-form $\X(\cL)$ which is constructed in terms of  
 $\cL$ and possesses the following properties: 
 (i) $\X(\cL)$ is closed, $\rd \, \X(\cL) =0$; 
(ii) $\X(\cL)$ is  super-Weyl invariant, $\d_\s \X(\cL)=0$; (iii) $\X(\cL)$ is dimensionless. 
In fact, the conditions (ii) and (iii) prove to completely 
determine (modulo an overall factor) the explicit form of 
$\X(\cL)$  to be
\bea
\X(\cL) &=&
\frac{\ri}{2}E^\g E^\b E^a
(\g_a)_{\b\g}\cL
+\frac{1}{4}E^\g E^b E^a\ve_{abc}
(\g^c)_{\g}{}^\d\cD_{\d}\cL
\non\\
&&
-\frac{1}{24}E^cE^bE^a\ve_{abc}
\big(\ri\cD^2+8\cS\big)\cL~.
\label{2.21}
\eea
It is a simple exercise to check that this form is closed, $\rd \,\X(\cL) =0$. 
This three-form coincides with  that originally constructed in  \cite{Becker:2003wb}
by directly solving the cohomological problem  $\rd \,\X=0$.


\section{Traditional approach to conformal supergravity and topologically massive supergravity}
\setcounter{equation}{0}

With the matrix notation $\O_A = (\O_{A\, \b}{}^\g )$, the action for  conformal supergravity is 
\bea
S_{\text{CSG}} &=& \int \rd^3x \rd^2 \q E\, 
\O^{\a\b\g} \Big\{ C_{\a\b\g} -\frac{4}{3} \ve_{\a(\b} \cD_{\g)} \cS\Big\} \non \\
&&- \frac{1}{3} \int \rd^3x \rd^2 \q E\,  \Big\{ \tr (\frac{1}{4} \O^\a \O^\b \O_{\a\b} +2 \cS \O^\a \O_\a) 
+ \cS \O^{\a \b}{}_\a \O^\g{}_{\b \g} \Big\} \non \\
&&+ 16\ri \int \rd^3x \rd^2 \q E\, \cS^2~.
\label{3.1}
\eea
The tensor in the braces in the first line of \eqref{3.1} is divergenceless 
with respect to its first index, 
\bea
\cD^\a \Big\{ C_{\a\b\g} -\frac{4}{3} \ve_{\a(\b} \cD_{\g)} \cS \Big\} =0~.
\eea
Modulo an overall coefficient, the structures in the first and second lines of \eqref{3.1} are uniquely 
fixed by the condition of invariance under the local Lorentz transformations 
\bea
\d_K \O_A{}^{bc} = K_A{}^D \O_D{}^{bc} - \cD_A K^{bc}~.
\eea
The last term in \eqref{3.1} is uniquely fixed by requiring invariance under the super-Weyl transformations 
\begin{subequations}
\bea
\d_\s \O_\a{}^{\g \d} &=& \hf \s \O_\a{}^{\g \d } + 2\d^{(\g}_\a \cD^{\d)} \s ~, \\
\d_\s \O_{\a\b}{}^{\g \d} &=&  \s \O_{\a\b}{}^{\g \d} - \ri (\cD_{(\a} \s) \O_{\b)}{}^{\g\d} 
+2\d^{(\g}_{(\a} \cD^{\d)}_{\b)} \s~.
\eea
\end{subequations}

The topological mass term appearing in the first two lines of \eqref{3.1} 
was given in {\it Superspace} \cite{GGRS}, eq. (2.6.47). 
However, since 3D $\cN=1$ super-Weyl invariance was not discussed in \cite{GGRS}, 
the complete action \eqref{3.1} was not presented in this book. 
Instead the requirement of super-Weyl invariance was put forward by Zupnik and Pak \cite{ZP88,ZP89}, 
who derived an action of the form \eqref{3.1}. 
However, their completely symmetric torsion $ C_{\a\b\g} $
was chosen to obey the constraint $\cD^\a  C_{\a\b\g} = 0$ which holds only in a special 
super-Weyl gauge such that $\cD_{\a\b} \cS=0$, due to the Bianchi identity \eqref{2.9}.
So the actual status of the action presented in \cite{ZP88,ZP89} is not quite clear to us. 

Varying the conformal supergravity action \eqref{3.1} with respect to a supergravity prepotential 
\cite{GGRS}  leads to the $\cN=1$ supersymmetric Cotton tensor\footnote{This expression
may be compared with the $\cN=2$ supersymmetric Cotton tensor \cite{K12}.} 
\bea
W_{\a\b\g} = \frac{\ri }{2}\, \cD^2 C_{\a\b \g} +  \cD_{(\a\b} \cD_{\g)} \cS + 4 \cS C_{\a\b \g} ~.
\eea
It is an instructive exercise to prove that the super-Weyl transformation of $W_{\a\b\g}$ is 
\bea
\d_\s W_{\a\b\g} = \frac{5}{2} \s W_{\a\b\g}~.
\label{3.6}
\eea
The equation of motion for conformal supergravity is 
\bea
W_{\a\b\g}=0~.
\label{3.6-7}
\eea 
At the same time, this equation
is the necessary and sufficient condition for the superspace to be conformally flat. 

The action for three-dimensional $\cN=1$ Poincar\'e supergravity is
\bea
S_{\rm PSG} = - \ri \int \rd^3x \rd^2 \q E\, \Big\{ 
\frac{\ri}{2} \cD^\a\varphi\cD_\a\varphi -\cS \varphi^2 
+ \frac{\l}{4}   \varphi^4 \Big\} ~,
\label{3.7}
\eea
with $\l$ a cosmological constant. This action is super-Weyl invariant provided $\vf$ transforms 
by the rule
\bea
\d_\s \vf = \hf \s \vf~.
\eea
The scalar field $\vf$, which is chosen to be nowhere vanishing,
plays the role of the conformal compensator. The super-Weyl invariance can be used to 
impose the gauge  $\vf =1$. 
 
The equations of motion corresponding to the action \eqref{3.7} are 
\begin{subequations}
\bea
\vf \cD_{(\a} \cD_{\b \g)} \vf  - 3 \cD_{(\a}\vf \cD_{\b \g )}\vf + \vf^2 C_{\a\b\g} &=&0~, 
\label{3.9a} \\
\big( {\ri }\, \cD^2 +2\cS \big)\vf  -\l \vf^3 &=&0~. 
\label{3.9b}
\eea
\end{subequations}
Eq. \eqref{3.9a} is obtained by varying the action with respect to the supergravity prepotential.
The left-hand side of \eqref{3.9a} is proportional to the supercurrent for the scalar superfield $\vf$ 
described by the action \eqref{3.7}.  
It is an instructive exercise to show that both equations \eqref{3.9a}  and \eqref{3.9b} are super-Weyl 
invariant. 

To describe topologically massive supergravity, we have to consider a linear combination 
of the two actions \eqref{3.1} and \eqref{3.7}. Then the conformal supergravity equation 
\eqref{3.6-7} turns into 
\bea
W_{\a\b\g} - m \Big( \vf \cD_{(\a} \cD_{\b \g)} \vf  - 3 \cD_{(\a}\vf \cD_{\b \g )}\vf 
+ \vf^2 C_{\a\b\g} \Big)=0~, 
\label{3.11}
\eea
with $m$ a parameter of unit mass dimension. 

If we choose the super-Weyl gauge $\vf =1$ and linearize eq. \eqref{3.11}
around Minkowski superspace, we end up with 
\bea
\Big(\frac{\ri }{2}\,D^2   -m \Big)C_{\a\b \g} =0 \quad \Longrightarrow \quad
(\Box -m^2 ) C_{\a\b\g} =0~,
\eea
with $D^2= D^\a D_\a$ and $D_\a$ the spinor derivatives in Minkowski superspace. 


\section{Main construction}
\setcounter{equation}{0}

In this section we describe the key points of our method as applied to 
$\cN=1$ conformal supergravity.

\subsection{One-parameter family of covariant derivatives}

Of special importance for our analysis is the observation that the geometry of conformal supergravity,
which was reviewed in section 2,   
can be described by a one-parameter family of covariant derivatives $\nabla_A$, 
\bea
\cD_A = E_A +\O_A \quad \longrightarrow \quad \de_A= E_A+{\bm \O}_A~,
\eea
defined as follows
\bsubeq
\bea
&\de_\a=\cD_\a~,\qquad
\de_{a}=\cD_a+2\l\cS\cM_a~,~~~
\label{dvcd}
\eea
with $\l$ a real parameter. Only the vector connection becomes deformed,
\bea
{\bm\O}_\a{}^{c}={\O}_\a{}^{c} ~, \qquad
{\bm\O}_a{}^{c}={\O}_a{}^{c}-2\l\cS\d_a{}^c ~.
\eea
\esubeq

We will use boldface notation for the torsion and curvature tensors 
associated with the deformed covariant derivatives $\nabla_A$,  
\bea
[\de_A,\de_B\}&=&{\bm T}_{AB}{}^C\de_C+\hf {\bm R}_{AB}{}^{cd}\cM_{cd}
~.
\eea
The algebra of the deformed covariant derivatives is as follows:
\bsubeq
\bea
\{\de_\a,\de_\b\}&=&
2\ri\de_{\a\b}
-4\ri(1+\l) \cS(\g^c)_{\a\b}\cM_c
~,
\label{de-N=1alg-1}
\\
{[}\de_a,\de_\b{]}
&=&
(1-\l)\cS(\g_a)_\b{}^{\g}\de_{\g}
-\Big{[}
(\g_a)_\b{}^{\g}(\g^c)^{\d\r}C_{\g\d\r}
+\frac{2(1+3\l)}{3}(\de_{\b}\cS)\d_a^c
\non\\
&&~~~~~~~~~~~~~~~~~~~~~~~~~~~~~
-\frac{4}{3}\ve_{ab}{}^{c}(\g^b)_{\b\g}(\de^{\g}\cS)
\Big{]}\cM_c
~,
\label{de-N=1alg-3/2}
\\
{[}\de_a,\de_b{]}
&=&
4\l\cS\ve_{ab}{}^{c}\de_{c}
+\ve_{abc}\Big{[}\frac{\ri}{2}(\g^c)_{\a\b}C^{\a\b\g}\de_\g
-\frac{2\ri}{3}(\g^c)^{\b\g}(\de_{\b}\cS)\Big{]}\de_\g
\non\\
&&
-\ve_{abc}\Big{[}
-\frac{\ri}{2}(\g^c)^{\a\b}(\g^d)^{\g\d}\de_{(\a}C_{\b\g\d)}
-\eta^{cd}\Big(\frac{2\ri}{3}(\de^2\cS)+4(1-\l^2)\cS^2\Big)
\non\\
&&~~~~~~~~~
+2\l\ve^{ced}(\de_{e}\cS)
\Big{]}\cM_{d}
~.
\label{de-N=1alg-2}
\eea
\esubeq
The covariant derivatives $\cD_A$, which were introduced in section 2,  
correspond to the choice $\l=0$. On the other hand, the choice $\l=-1$ 
corresponds to the
covariant derivatives employed in the book \cite{GGRS}.

The spinor covariant derivative 
$\de_\a$ and the torsion tensors $\cS$ and $C_{\a\b\g}$ are obviously $\l$-independent, 
and their super-Weyl transformations do not change.  
The super-Weyl transformation of 
the deformed vector derivatives $\de_a$ is
\bea
\d_\s\de_a
&=&
\s\de_a
+{\ri\over 2}(\g_a)^{\g\d}(\de_{\g} \s)\de_{\d}
+\ve_{abc}(\de^b\s)\cM^{c}
-\frac{\l\ri}{2}(\de^2\s)\cM_a
~.
\eea
${}$From here we read off the super-Weyl transformation 
of the deformed connection one-form:
\bea
\d_\s{\bm \O}{}^{c}&=&
E^\a\Big\{-(\g^c)_{\a\b}(\de^{\b}\s)\Big\}
+E^{a}\Big\{-\ve_{ab}{}^c(\de^{b}\s)
+\frac{\l\ri}{2}\d_a^c(\de^2\s)\Big\}
~.
\eea
This turns into \eqref{2.18} for $\l=0$.

\subsection{Curvature squared  four-form}

We are interested in the $\l$-dependent four-form
\bea
{\bm R}^2:={\bm R}^a  {\bm R}_a=\frac{2}{5}{\bm R}_A{}^B  {\bm R}_B{}^A~,
\eea
which is closed for any value of $\l$,
\bea
\rd {\bm R}^2=0
~.
\eea
The curvature two-form 
${\bm R}^d= \hf E^B E^A {\bm R}_{AB}{}^d$
has the following components:
\bsubeq
\bea
{\bm R}_{\a\b}{}^d&=&4\ri(1+\l)\cS(\g^d)_{\a\b}
~,
\\
{\bm R}_{a\b}{}^d&=&
(\g_a)_\b{}^{\g}(\g^d)^{\d\r}C_{\g\d\r}
+\frac{1}{3}\big(
2(1+3\l)(\de_{\b}\cS)\d_a^d
-4\ve_{ab}{}^{d}(\g^b)_{\b\g}(\de^{\g}\cS)\big)
~,
\\
{\bm R}_{ab}{}^d&=&\ve_{abc}\Big{[}
-\frac{\ri}{2}(\g^c)^{\a\b}(\g^d)^{\g\d}\de_{(\a}C_{\b\g\d)}
-\Big(\frac{2\ri}{3}(\de^2\cS)+4(1-\l^2)\cS^2\Big)\eta^{cd}
\non\\
&&~~~~~~
+2\l\ve^{ced}(\de_{e}\cS)\Big{]}
~.
\eea
\esubeq
In terms of the  curvature two-form, 
the form ${\bm R}^2$ reads 
\bea
{\bm R}^2&:=&{\bm R}^e  {\bm R}_e=
\frac{1}{4}E^DE^CE^BE^A{\bm R}_{AB}{}^e{\bm R}_{CD}{}_e
\non\\
&=&
\frac{1}{4}E^\d E^\g E^\b E^\a {\bm R}_{\a\b}{}^e{\bm R}_{\g\d}{}_e
+E^\d E^\g E^\b E^a{\bm R}_{a\b}{}^e{\bm R}_{\g\d}{}_e
\non\\
&&
+\frac{1}{2}E^\d E^\g E^bE^a {\bm R}_{ab}{}^e{\bm R}_{\g\d}{}_e
-E^\d E^\g E^bE^a{\bm R}_{a\g}{}^e{\bm R}_{b\d}{}_e
\non\\
&&
+E^\d E^cE^bE^a{\bm R}_{ab}{}^e{\bm R}_{c\d}{}_e
~.
\eea
Using this representation and also defining the components of ${\bm R}^2$ as
\bea
{\bm R}^2
&=&\frac{1}{4!}E^DE^CE^BE^A({\bm R}^2)_{ABCD}
\non\\
&=&
\frac{1}{24}E^\d E^\g E^\b E^\a ({\bm R}^2)_{\a\b\g\d}
+\frac{1}{6}E^\d E^\g E^\b E^a({\bm R}^2)_{a\b\g\d}
+\frac{1}{4}E^\d E^\g E^bE^a({\bm R}^2)_{ab\g\d}
\non\\
&&
+\frac{1}{6}E^\d E^c E^b E^a({\bm R}^2)_{abc \d}
~,
\eea
direct calculations give
\bsubeq
\bea
({\bm R}^2)_{\a\b\g\d}&=&0
~,
\\
({\bm R}^2)_{a\b\g\d}&=&
-48\ri(1+\l)\cS(\g_a)_{(\b}{}^{\r}C_{\g\d)\r}
-16\ri(1+\l)(1-3\l)\cS(\de_{(\b}\cS)(\g_a)_{\g\d)}~,
\\
({\bm R}^2)_{ab\g\d}&=&
\ve_{abc}\Big\{\,
4(\g^c)_{\g\d}C^{\a\r\t}C_{\a\r\t}
-8(\g^c)^{\r\t}(\de_{(\r}\cS)C_{\g\d)\t}
-\frac{8}{3}(\g^c)_{(\g}{}^{\r}(\de^{\t}\cS)C_{\d)\r\t}
\non\\
&&~~~~~~
-8\l(\g^c)^{\r\t}(\de_{(\g}\cS)C_{\d)\r\t}
-\frac{32\l}{3}(\g^c)_{\g\d}(\de^{\r}\cS)(\de_{\r}\cS)
\non\\
&&~~~~~~
-8(1+\l)\cS\Big{[}(\g^c)^{\r\t}(\de_{(\g}C_{\d\r\t)})
+\ri(\g^c)_{\g\d}\Big(\frac{2\ri}{3}(\de^2\cS)+4(1-\l^2)\cS^2\Big)
\non\\
&&~~~~~~~~~~~~~~~~~~~~~~~~
+2\ri\l\ve^{cde}(\g_d)_{\g\d}(\de_{e}\cS)\cS
\Big{]}
\Big\}
~,
\\
({\bm R}^2)_{abc\d}&=&
\ve_{abc}\Big\{
4\ri C^{\g\r\t}\de_{(\d}C_{\g\r\t)}
-8(1+3\l)\Big{[}\frac{\ri}{3}(\de^2\cS)+2(1-\l^2)\cS^2\Big{]}(\de_{\d}\cS)
\non\\
&&~~~~~~
-4\l(\g^d)^{\r\t}(\de_{d}\cS)C_{\d\r\tau}
+\frac{32\l}{3}(\g^d)_{\d\r}(\de^{\r}\cS)(\de_{d}\cS)
\Big\}
~.
\eea
\esubeq

\subsection{Torsion-induced three-form}

Our next task is to look 
for a three-form ${\bm \S}_{\rm T}=\frac{1}{3!}E^C  E^B  E^A{\bm \S}_{ABC}$  such that 
 (i) it obeys the equation
\bea
\rd {\bm \S}= {\bm R}^2~;
\label{DO3R2}
\eea
and 
(ii) its components are constructed in terms of the torsion and its covariant derivatives.
The equation \eqref{DO3R2} is equivalent to
\bea
4\de_{[A}{\bm \S}_{BCD)}-6{\bm T}_{[AB}{}^E{\bm\S}_{|E|CD)}=({\bm R}^2)_{ABCD}
~.
\label{DOR2}
\eea
Under the additional conditions ${\bm \S}_{\a\b\g}={\bm \S}_{a\b\g}=0$, 
it turns out that eq. \eqref{DOR2} allows us 
to completely determine all the components of 
${\bm \S}_{ABC}$
in terms of $({\bm R}^2)_{ABCD}$. 
This relies on the fact that
$T_{\a\b}{}^c=2\ri(\g^c)_{\a\b}$. The presence of this dimensionless torsion allows us to iteratively  
use \eqref{DOR2} to express ${\bm \S}_{ABC}$ in terms of derivatives of lower mass dimension 
components and $({\bm R}^2)_{ABCD}$.
This is a superform analogue of Dragon's theorem \cite{Dragon}
(see also \cite{Novak})
and is a crucial ingredient in iteratively
solving superspace Bianchi identities.
In particular, by using \eqref{DOR2}, it is not difficult to prove  that
\bsubeq
\bea
{\bm \S}_{ab\g}&=&\frac{\ri}{40}\ve_{abc} \Big{[}(\g^c)_\g{}^\d(\g^d)^{\r\t}({\bm R}^2)_{d\d\r\t}
+2\ve^{cde}(\g_d)^{\d\r}({\bm R}^2)_{e\g\d\r}\Big{]}
~,
\\
{\bm \S}_{abc}&=&-\frac{\ri}{24}\ve_{abc}(\g_d)^{\g\d}\ve^{def}\Big{[}
2\de_{(\g}{\bm \S}_{ef\d)}
-({\bm R}^2)_{ef\g\d}
\Big{]}
~.~~~~~~
\eea
\esubeq
Using the explicit form for $({\bm R}^2)_{ABCD}$ gives 
\bea
{\bm \S}_{\rm T}&=&\frac{1}{2}E^\g  E^b  E^a\,\ve_{abc}\Big\{
-4(1+\l)(\g^c)^{\d\r}C_{\g\d\r}\cS
-\frac{8(1+\l)(1-3\l)}{3}(\g^c)_{\g}{}^\d(\de_{\d}\cS)\cS
\Big\}
\non\\
&&
+\frac{1}{6}E^c  E^b  E^a\,\ve_{abc}\Big\{\,
2\ri C^{\a\b\g}C_{\a\b\g}
+\frac{4\ri}{3}(1-6\l-3\l^2)(\de^{\g}\cS)\de_{\g}\cS
\non\\
&&~~~~~~~~~~~~~~~~~~~~~~
+4\ri(1+\l)(1-\l)(\de^2\cS)\cS
+16(1+\l)(1-\l^2)\cS^3
\Big\}
~.~~~~~~~~~
\eea
One can explicitly check that  this three-form satisfies \eqref{DO3R2} 
or equivalently  \eqref{DOR2}.
The crucial feature of ${\bm \S}_{\rm T}$ is that it is constructed only in terms of 
the torsion tensor.

There is a natural freedom in the choice of ${\bm \S}_{\rm T}$ described by 
\bea
{\bm \S}_{\rm T} ~\to ~  \widetilde{\bm \S}_{\rm T} 
= {\bm \S}_{\rm T} + A\, \X( S^2)~,
\eea
where $A$ is a real parameter, and  $\X(\cL)$ denotes the closed three-form \eqref{2.21}.
The specific feature of the three-form $ \widetilde{\bm \S}_{\rm T}$ with  $A\neq 0$
is that $\widetilde{\bm \S}_{a\b\g} \neq 0$.

We need to work out
how ${\bm \S}_{\rm T}$ behaves under the super-Weyl transformations. For this 
it is necessary to use 
 the transformation rules for the supervielbein  \eqref{sWE-1/2}--\eqref{sWE-1}
and the torsion superfields \eqref{sWS}--\eqref{sWC} along with the following 
equations
\bsubeq
\bea
\de_\a\de^2\s&=&
-2\ri\de_{\a\b}\de^\b\s
+2\ri(1+3\l)\cS\de_\a\s
=
-2\ri\de^\b\de_{\a\b}\s
+8\ri\cS\de_\a\s
~,
\\
\de^2\de^2\s&=&
-4\de^{a}\de_{a}\s
+8\ri(\de^\a\cS)\de_\a\s
+8\ri\cS\de^2\s
~.
\eea
\esubeq
After some involved algebra,  we end up with 
 the super-Weyl variation of ${\bm \S}_{\rm T}$:
\bea
\d_\s{\bm \S}_{\rm T}&=&
\frac{1}{2}E^\g E^b E^a\ve_{abc}\Big\{
-4(1+\l)^2(1-3\l)(\g^c)_{\g}{}^\d\cS^2\de_\d\s
+(1+\l)\ri(\g^c)^{\d\r}C_{\g\d\r}\de^2\s
\non\\
&&~~~~~~~~~~~~~~~~~~~
+\frac{2\ri(1+\l)(1-3\l)}{3}(\g^c)_{\g}{}^\d(\de_{\d}\cS)\de^2\s
-4(1+\l)(1-\l)\cS\de^c\de_\g\s
\non\\
&&~~~~~~~~~~~~~~~~~~~
-4\l(1+\l)\ve^{cde}(\g_d)_{\g}{}^{\d}\cS\de_{e}\de_{\d}\s
\Big\}
\non\\
&&
+\frac{1}{6}E^c E^b E^a\ve_{abc}\Big\{
-2\ri(\g^d)_{\a\b} C^{\a\b\g}\de_{d}\de_{\g}\s
-\frac{4\ri(1-6\l-3\l^2)}{3}(\g^d)^{\g\d}(\de_{\g}\cS)\de_d\de_\d\s
\non\\
&&~~~~~~~~~~~~~~~~~~~~~~
+(1+\l)(1-\l)\Big(
-12\l\ri\cS^2\de^2\s
+(\de^2\cS)\de^2\s
-4\cS\de^{a}\de_{a}\s\Big)
\non\\
&&~~~~~~~~~~~~~~~~~~~~~~
+12(1+\l)(1-2\l-\l^2)\ri\cS(\de^{\a}\cS)\de_\a\s
\Big\}
~.
\label{sWO-0}
\eea
This variation is non-zero for any value of $\l$.


\subsection{Chern-Simons three-form}

We have just constructed the torsion-induced three-form ${\bm \S}_{\rm T}$
which solves the equation \eqref{DO3R2}.
The same equation has another natural solution given by a  
Lorentz Chern-Simons three-form defined by
\bea
{\bm \S}_{\rm CS}&:=&
{\bm \O}^c  {\bm R}_c
+ \frac{1}{6}\ve_{abc}{\bm \O}^{a} {\bm \O}^b {\bm \O}^{c}
=
\frac{2}{5}\Big{[}{\bm \O}_A{}^{B}  {\bm R}_B{}^A
+ \frac{1}{3}{\bm \O}_A{}^B {\bm \O}_B{}^C {\bm \O}_C{}^A
\Big{]}
~.~~~~~
\eea
Indeed, using the structure equation
\bea
\rd {\bm R}_A{}^B={\bm R}_A{}^C  {\bm \O}_C{}^B-{\bm \O}_A{}^C  {\bm R}_C{}^B
~,
\eea
it is easy to check that
\bea
\rd {\bm \S}_{\rm CS}={\bm R}^2
~.
\eea
There are two main differences between the three-forms 
${\bm \S}_{\rm T}$ and $ {\bm \S}_{\text CS}$. 
First of all, the components of  
${\bm \S}_{\rm T}$ are tensors constructed in terms of the torsion and its covariant derivatives, while  
the components of 
the Chern-Simons three-form $ {\bm \S}_{\rm CS}$ involve the naked connection.
Secondly, the three-form ${\bm \S}_{\rm T}$ is invariant under the local Lorentz transformations, 
while $ {\bm \S}_{\rm CS}$ changes by an exact term. 

The explicit form of $ {\bm \S}_{\rm CS}$ 
is as follows:
\bea
{\bm \S}_{\rm CS}&=&
\frac{1}{6}E^\g E^\b E^\a \Big\{
12\ri(1+\l)(\g^a)_{\a\b}\cS{\bm \O}_\g{}_a
- \ve_{abc}{\bm \O}_\a{}^a{\bm \O}_\b{}^b{\bm \O}_\g{}^c
\Big\}
\non\\
&&
+\frac{1}{2}E^\g E^\b E^a\Big\{
\Big{[}
2(\g_a)_\b{}^{\a}(\g^b)^{\d\r}C_{\a\d\r}
+\frac{4}{3}\Big{(}
(1+3\l)(\de_{\b}\cS)\d_a^c
+2\ve_{ab}{}^{c}(\g^b)_{\b}{}^{\a}(\de_{\a}\cS)\Big)
\Big{]}{\bm \O}_\g{}_c
\non\\
&&~~~~~~~~~~~~~~~~~
+4\ri(1+\l)(\g^c)_{\b\g}\cS{\bm \O}_a{}_c 
-\ve_{bcd}{\bm \O}_a{}^b {\bm \O}_\b{}^c {\bm \O}_\g{}^d
\Big\}
\non\\
&&
+\frac{1}{2}E^\g E^b E^a\ve_{abc}
\Big\{
\Big{[}-\frac{\ri}{2}(\g^c)^{\a\b}(\g^d)^{\d\r}\de_{(\a}C_{\b\d\r)}
-\eta^{cd}\Big(\frac{2\ri}{3}(\de^2\cS)+4(1-\l^2)\cS^2\Big)
\non\\
&&~~~~~~~~~~~~~~~~~~~~~~~~
+2\l\ve^{ced}(\de_{e}\cS)
\Big{]}{\bm \O}_\g{}_d
\non\\
&&~~~~~~~~~~~~~~~~~~~~~
+\Big{[}
\ve^{cdf}(\g_{f})_{\g\a}(\g^e)_{\d\r}C^{\a\d\r}
-\frac{2(1+3\l)}{3}(\de_{\g}\cS)\ve^{cde}
\non\\
&&~~~~~~~~~~~~~~~~~~~~~~~~~~
-4\eta^{e[c}(\g^{d]})_{\g}{}^{\a}(\de_{\a}\cS)
\Big{]}{\bm \O}_{de} 
\non\\
&&~~~~~~~~~~~~~~~~~~~~~
+\hf\ve^{cd_1d_2} \ve_{e_1e_2e_3}{\bm \O}_{d_1}{}^{e_1} {\bm \O}_{d_2}{}^{e_2} {\bm \O}_\g{}^{e_3}
\Big\}
\non\\
&&
+\frac{1}{6}E^cE^bE^a\ve_{abc}\Big\{
\Big{[}-\frac{\ri}{2}(\g^d)^{\a\b}(\g^e)^{\g\d}\de_{(\a}C_{\b\g\d)}
-\eta^{de}\Big(\frac{2\ri}{3}(\de^2\cS)+4(1-\l^2)\cS^2\Big)
\non\\
&&~~~~~~~~~~~~~~~~~~~~~~~
-2\l\ve^{def}(\de_{f}\cS)
\Big{]}{\bm \O}_{de} 
\non\\
&&~~~~~~~~~~~~~~~~~~~~~
+\frac{1}{6}\ve_{abc}\ve^{d_1d_2d_3}\ve_{e_1e_2e_3}
{\bm \O}_{d_1}{}^{e_1} {\bm \O}_{d_2}{}^{e_2} {\bm \O}_{d_3}{}^{e_3} 
\Big\}
~.
\eea

We conclude this subsection by giving the expression for the super-Weyl transformation 
of ${\bm \S}_{\rm CS}$. 
In computing its variation, we can ignore all contributions that are exact three-forms. 
This considerably simplifies the calculation if we make use of the identity
\bea
\d_\s {\bm \S}_{\rm CS}&=&2(\d_\s{\bm \O}^a)  {\bm R}_a+
\mbox{exact three-form}
~.
\eea
The result of the calculation is
\bea
\d_\s{\bm \S}_{\rm CS}&=&
\frac{1}{2}E^\g E^\b E^a\Big\{
-4\ve_{abc}(\g^b)_{\b\g}(\g^c)_{\r\a}C^{\r\a\d}\de_{\d}\s
+\frac{4(1+3\l)}{3}\ve_{abc}(\g^b)_{\b\g}(\g^c)^{\r\a}(\de_{\a}\cS)\de_{\r}\s
\non\\
&&~~~~~~~~~~~~~~~
-4(1-\l)(\g_a)_{\b\g}(\de^{\a}\cS)\de_{\a}\s
-4\l(1+\l)(\g_a)_{\b\g}\cS\de^2\s
\non\\
&&~~~~~~~~~~~~~~~
+8\ri(1+\l)\ve_{abc}(\g^b)_{\b\g}\cS\de^{c}\s
\Big\}
\non\\
&&
+\frac{1}{2}E^\g E^b E^a
\ve_{abc}\Big\{
-2\ri(\g^c)^{\a\b}(\de_{(\a}C_{\b\g\d)})\de^{\d}\s
+\l\ri(\g^c)^{\d\r}C_{\g\d\r}\de^2\s
\non\\
&&~~~~~~~~~~~~~~~~~~~~~~
+2(\g_{d})_{\g\a}(\g^c)_{\d\r}C^{\a\d\r}\de^{d}\s
-(\g^c)_{\g}{}^{\r}\Big{[}\frac{4\ri}{3}(\de^2\cS)+8(1-\l^2)\cS^2\Big{]}\de_{\r}\s
\non\\
&&~~~~~~~~~~~~~~~~~~~~~~
-4\l\ve^{cde}(\g_d)_{\g}{}^{\r}(\de_{e}\cS)\de_{\r}\s
-\frac{8\l\ri}{3}(\g^c)_{\g\a}(\de^{\a}\cS)\de^2\s
\non\\
&&~~~~~~~~~~~~~~~~~~~~~~
+\frac{8}{3}(1+3\l)(\de_{\g}\cS)\de^{c}\s
-\frac{8}{3}\ve^{cde}(\g_d)_{\g\a}(\de^{\a}\cS)\de_e\s
\Big\}
\non\\
&&
+\frac{1}{6}E^cE^bE^a\ve_{abc}\Big\{
8\l(\de_{e}\cS)\de^{e}\s
+2\l(\de^2\cS)\de^2\s
-12\ri\l(1+\l)(1-\l)\cS^2\de^2\s
\Big\}
\non\\
&&
+\mbox{exact three-form}
~.
\label{sWCS-0}
\eea


\subsection{Closed three-forms}

An immediate corollary of
the results obtained so far is that the three-form defined by
\bea
{\bm \S} :={\bm \S}_{\rm T}-{\bm \S}_{\rm CS}
\eea
is closed,
\bea
\rd{\bm \S} =0
~.
\eea
Now one can see the advantage of using the deformed covariant derivatives $\nabla_A$, 
eq. \eqref{dvcd}, which depend on the parameter $\l$.
The crucial observation  is that 
the three-form 
${\bm \S} $ is closed for any value of $\l$, 
and so are  its partial derivatives with
respect to $\l$.
Because ${\bm \S} $ is polynomial in $\l$,  
by differentiating  ${\bm \S} $ with respect to $\l$
we  generate a finite number of new closed three-forms.
In particular, the explicit form of ${\bm \S} $ is
\bea
{\bm \S} &=&
\frac{1}{6}E^\g E^\b E^\a \Big\{
-12\ri(1+\l)(\g^a)_{\a\b}\cS{\bm \O}_\g{}_a
+ \ve_{abc}{\bm \O}_\a{}^a{\bm \O}_\b{}^b{\bm \O}_\g{}^c
\Big\}
\non\\
&&
+\frac{1}{2}E^\g E^\b E^a\Big\{
\Big{[}\,
\frac{8}{3}\ve_{ab}{}^{c}(\g^b)_{\b\a}(\de^{\a}\cS)
-\frac{4(1+3\l)}{3}(\de_{\b}\cS)\d_a^c
-2(\g_a)_\b{}^{\a}(\g^c)^{\d\r}C_{\a\d\r}
\Big{]}{\bm \O}_\g{}_c
\non\\
&&~~~~~~~~~~~~~~~~~
-4\ri(1+\l)(\g^c)_{\b\g}\cS{\bm \O}_a{}_c 
+\ve_{bcd}{\bm \O}_a{}^b {\bm \O}_\b{}^c {\bm \O}_\g{}^d
\Big\}
\non\\
&&
+\frac{1}{2}E^\g E^b E^a\ve_{abc}
\Big\{
-4(1+\l)(\g^c)^{\d\r}C_{\g\d\r}\cS
-\frac{8(1+\l)(1-3\l)}{3}(\g^c)_{\g}{}^\d(\de_{\d}\cS)\cS
\non\\
&&~~~~~~~~~~~~~~~~~~~~~
+\Big{[}\frac{\ri}{2}(\g^c)^{\a\b}(\g^e)^{\r\d}\de_{(\a}C_{\b\r\d)}
+\Big(\frac{2\ri}{3}(\de^2\cS)+4(1-\l^2)\cS^2\Big)\eta^{ce}
\non\\
&&~~~~~~~~~~~~~~~~~~~~~~~~~~
-2\l\ve^{cde}(\de_{d}\cS)
\Big{]}{\bm \O}_\g{}_e
\non\\
&&~~~~~~~~~~~~~~~~~~~~~
+\Big{[}
\ve^{cde}(\g_{e})_\g{}^{\a}(\g^f)^{\d\r}C_{\a\d\r}
+\frac{2(1+3\l)}{3}(\de_{\g}\cS)\ve^{cdf}
\non\\
&&~~~~~~~~~~~~~~~~~~~~~~~~~~
+\frac{8}{3}\eta^{f[c}(\g^{d]})_{\g}{}^{\a}(\de_{\a}\cS)
\Big{]}{\bm \O}_{d}{}_f 
\non\\
&&~~~~~~~~~~~~~~~~~~~~~
-\frac{1}{2}\ve^{cde} \ve_{fgh}{\bm \O}_d{}^f {\bm \O}_e{}^g {\bm \O}_\g{}^h
\Big\}
\non\\
&&
+\frac{1}{6}E^cE^bE^a\ve_{abc}\Big\{
2\ri C^{\mu\r\t}C_{\mu\r\t}
+\frac{4\ri(1-6\l-3\l^2)}{3}(\de^{\g}\cS)(\de_{\g}\cS)
\non\\
&&~~~~~~~~~~~~~~~~~~~~~
+4\ri(1+\l)(1-\l)(\de^2\cS)\cS
+16(1+\l)(1-\l^2)\cS^3
\non\\
&&~~~~~~~~~~~~~~~~~~~~~
+\Big{[}\frac{\ri}{2}(\g^d)^{\a\b}(\g^f)^{\g\d}\de_{(\a}C_{\b\g\d)}
+\Big(\frac{2\ri}{3}(\de^2\cS)+4(1-\l^2)\cS^2\Big)\eta^{df}
\non\\
&&~~~~~~~~~~~~~~~~~~~~~~~~~~
-2\l\ve^{def}(\de_{e}\cS)
\Big{]}{\bm \O}_d{}_f 
\non\\
&&~~~~~~~~~~~~~~~~~~~~~
-\frac{1}{6}\ve^{e_1e_2e_3}\ve_{d_1 d_2 d_3}
{\bm \O}_{e_1}{}^{d_1} {\bm \O}_{e_2}{}^{d_2} {\bm \O}_{e_3}{}^{d_3} 
\Big\}~.
\label{4.28}
\eea
In computing partial derivatives of $\bm \S$ with respect to $\l$, one has to use the identities
\bea
\pa_\l{\bm\O}_a{}^c =-2\d_a^c\cS~,~~~~~~
\pa_\l\de_a=2\cS\cM_a
~.
\eea
Direct calculations give
\bea
\frac{\pa{\bm \S}}{\pa\l}&=&
\frac{1}{6}E^\g E^\b E^\a \Big\{
-12\ri(\g^a)_{\a\b}\cS{\bm \O}_\g{}_a
\Big\}
\non\\
&&
+\frac{1}{2}E^\g E^\b E^a\Big\{
8\ri(1+\l)(\g_a)_{\b\g}\cS^2
-4(\de_{\b}\cS){\bm \O}_\g{}_a
-4\ri(\g^c)_{\b\g}\cS{\bm \O}_a{}_c 
-2\ve_{acd}\cS{\bm \O}_\b{}^c {\bm \O}_\g{}^d
\Big\}
\non\\
&&
+\frac{1}{2}E^\g E^b E^a\ve_{abc}
\Big\{
-2(\g^c)^{\d\r}C_{\g\d\r}\cS
+\frac{16(2+3\l)}{3}(\g^c)_{\g}{}^\d(\de_{\d}\cS)\cS
+2\ve^{cdf}(\de_{\g}\cS){\bm \O}_{df} 
\non\\
&&~~~~~~~~~~~~~~~~~~~~~
+\Big{[}
-8\l\cS^2\eta^{ce}
-2\ve^{cde}(\de_{d}\cS)
\Big{]}{\bm \O}_\g{}_e
+4\cS{\bm \O}_e{}^{[c} {\bm \O}_\g{}^{e]}
\Big\}
\non\\
&&
+\frac{1}{6}E^cE^bE^a\ve_{abc}\Big\{
-8\ri(1+\l)(\de^{\g}\cS)(\de_{\g}\cS)
-4\ri(1+2\l)(\de^2\cS)\cS
\non\\
&&~~~~~~~~~~~~~~~~~~~~~
-8(1+\l)(1+3\l)\cS^3
+\Big{[}
-8\l\cS^2\eta^{df}
-2\ve^{def}(\de_{e}\cS)
\Big{]}{\bm \O}_d{}_f 
\non\\
&&~~~~~~~~~~~~~~~~~~~~~
-2\cS{\bm \O}_{d}{}^{[d} {\bm \O}_{e}{}^{e]} 
\Big\}
~,
\label{4.30}
\eea
as well as 
\bea
\frac{\pa^2{\bm \S} }{\pa\l^2}&=&
\frac{1}{2}E^\g E^\b E^a\Big\{
\ri(\g_a)_{\b\g}(16\cS^2)
\Big\}
+\frac{1}{2}E^\g E^b E^a\ve_{abc}
\Big\{
\hf(\g^c)_{\g}{}^\d\de_{\d}(16\cS^2)
\Big\}
\non\\
&&
+\frac{1}{6}E^cE^bE^a\ve_{abc}\Big\{
-\frac{1}{4}\big(\ri\de^2+8\cS\big)(16\cS^2)
\Big\}\non \\
&=& \X(16 \cS^2)
~,
\label{4.31}
\eea
with the closed three-form $\X (\cL)$ defined by \eqref{2.21}. Since the three-form 
\eqref{4.31} is $\l$-independent, we conclude that
\bea
\frac{\pa^3{\bm \S}}{\pa\l^3}=0
~.
\eea
Thus we have constructed the closed three-forms \eqref{4.28}, 
\eqref{4.30} and \eqref{4.31}.

The super-Weyl variations of ${\bm \S}$ and 
its first and second derivatives with respect to $\l$ can be  computed by representing 
$\d_\s{\bm \S}=(\d_\s{\bm \S}_{\rm T}-\d_\s{\bm \S}_{\rm CS})$
and then making use of
\eqref{sWO-0} and \eqref{sWCS-0}.
It holds that
\bea
\d_\s{\bm \S} &=&
\frac{1}{2}E^\g E^\b E^a\Big\{
4\ve_{abc}(\g^b)^{\b\g}(\g^c)_{\r\a}C^{\r\a\d}\de_{\d}\s
+4(1-\l)(\g_a)_{\b\g}(\de^{\a}\cS)\de_{\a}\s
\non\\
&&~~~~~~~~~~~~~~
-\frac{4(1+3\l)}{3}\ve_{abc}(\g^b)_{\b\g}(\g^c)^{\r\a}(\de_{\a}\cS)\de_{\r}\s
+4(\l+\l^2)(\g_a)_{\b\g}\cS\de^2\s
\non\\
&&~~~~~~~~~~~~~~
+8\ri(1+\l)\ve_{abc}(\g^c)_{\b\g}\cS\de^{b}\s
\Big\}
\non\\
&&
+\frac{1}{2}E^\g E^b E^a\ve_{abc}\Big\{
2\ri(\g^c)^{\a\b}(\de_{(\a}C_{\b\g\d)})\de^{\d}\s
+\ri(\g^c)^{\d\r}C_{\g\d\r}\de^2\s
\non\\
&&~~~~~~~~~~~~~~~~~~~~~
-2(\g^{d})_{\g\a}(\g^c)_{\d\r}C^{\a\d\r}\de_{d}\s
+4(1+\l+3\l^2+3\l^3)(\g^c)_{\g}{}^\d\cS^2\de_\d\s
\non\\
&&~~~~~~~~~~~~~~~~~~~~~
+4\l\ve^{cde}(\g_d)_{\g}{}^{\r}(\de_{e}\cS)\de_{\r}\s
+\frac{4\ri}{3}(\g^c)_{\g}{}^{\r}(\de^2\cS)\de_{\r}\s
\non\\
&&~~~~~~~~~~~~~~~~~~~~~
+\frac{2\ri(1-6\l-3\l^2)}{3}(\g^c)_{\g}{}^\d(\de_{\d}\cS)\de^2\s
-4(1-\l^2)\cS\de^c\de_\g\s
\non\\
&&~~~~~~~~~~~~~~~~~~~~~
-4(\l+\l^2)\ve^{cde}(\g_d)_{\g}{}^{\d}\cS\de_{e}\de_{\d}\s
-\frac{8}{3}(1+3\l)(\de_{\g}\cS)\de^{c}\s
\non\\
&&~~~~~~~~~~~~~~~~~~~~~
+\frac{8}{3}\ve^{cde}(\g_d)_{\g\a}(\de^{\a}\cS)\de_{e}\s
\Big\}
\non\\
&&
+\frac{1}{6}E^c E^b E^a\ve_{abc}\Big\{
-2\ri(\g^d)_{\a\r} C^{\a\r\t}\de_{d}\de_{\t}\s
+12\ri(1-\l-3\l^2-\l^3)\cS(\de^{\a}\cS)\de_\a\s
\non\\
&&~~~~~~~~~~~~~~~~~~~~~
+(1-2\l-\l^2)(\de^2\cS)\de^2\s
-4(1-\l^2)\cS\de^{e}\de_{e}\s
\non\\
&&~~~~~~~~~~~~~~~~~~~~~~
-8\l(\de^{e}\cS)\de_{e}\s
-\frac{4\ri(1-6\l-3\l^2)}{3}(\g^d)^{\g\d}(\de_{\g}\cS)\de_d\de_\d\s
\Big\}
\non\\
&&
+   \mbox{exact three-form}
~,
\label{sWvard0}
\eea
and
\bea
\d_\s\frac{\pa{\bm \S} }{\pa\l}&=&
\frac{1}{2}E^\g E^\b E^a\Big\{
-4(\g_a)_{\b\g}(\de^{\a}\cS)\de_{\a}\s
-4\ve_{abc}(\g^b)_{\b\g}(\g^c)^{\r\a}(\de_{\a}\cS)\de_{\r}\s
\non\\
&&~~~~~~~~~~~~~~~
+4(1+2\l)(\g_a)_{\b\g}\cS\de^2\s
+8\ri\ve_{abc}(\g^c)_{\b\g}\cS\de^{b}\s
\Big\}
\non\\
&&
+\frac{1}{2}E^\g E^b E^a\ve_{abc}\Big\{
8(1+2\l+3\l^2)(\g^c)_{\g}{}^\d\cS^2\de_\d\s
+4\ve^{cde}(\g_d)_{\g}{}^{\d}(\de_{e}\cS)\de_{\d}\s
\non\\
&&~~~~~~~~~~~~~~~~~~~~~
-4(1+\l)\ri(\g^c)_{\g}{}^\d(\de_{\d}\cS)\de^2\s
+8\l\cS\de^c\de_\g\s
-8(\de_{\g}\cS)\de^{c}\s
\non\\
&&~~~~~~~~~~~~~~~~~~~~~
-4(1+2\l)\ve^{cde}(\g_d)_{\g}{}^{\d}\cS\de_{e}\de_{\d}\s
\Big\}
\non\\
&&
+\frac{1}{6}E^c E^b E^a\ve_{abc}\Big\{
-8\ri(2+6\l+3\l^2)\cS(\de^{\a}\cS)\de_\a\s
-2(1+\l)(\de^2\cS)\de^2\s
\non\\
&&~~~~~~~~~~~~~~~~~~~~~
-8(\de^{e}\cS)\de_{e}\s
+8\ri(1+\l)(\g^d)^{\g\d}(\de_{\g}\cS)\de_d\de_\d\s
+8\l\cS\de^{e}\de_{e}\s
\Big\}
\non\\
&&
+  \mbox{exact three-form}
~,
\label{sWvard1}
\eea
and finally 
\bea
\d_\s\frac{\pa^2{\bm \S} }{\pa\l^2}&=&
\frac{1}{2}E^\g E^\b E^a\Big\{
8(\g_a)_{\b\g}\cS\de^2\s
\Big\}
\non\\
&&
+\frac{1}{2}E^\g E^b E^a\ve_{abc}\Big\{
8(1+3\l)(\g^c)_{\g}{}^\d\cS^2\de_\d\s
-4\ri(\g^c)_{\g}{}^\d(\de_{\d}\cS)\de^2\s
\non\\
&&~~~~~~~~~~~~~~~~~~~~~~
+8\cS\de^c\de_\g\s
-8\ve^{cde}(\g_d)_{\g}{}^{\d}\cS\de_{e}\de_{\d}\s
\Big\}
\non\\
&&
+\frac{1}{6}E^c E^b E^a\ve_{abc}\Big\{
-24\ri(1+\l)\cS(\de^{\a}\cS)\de_\a\s
-2(\de^2\cS)\de^2\s
\non\\
&&~~~~~~~~~~~~~~~~~~~~~~
+8\ri(\g^d)^{\g\d}(\de_{\g}\cS)\de_d\de_\d\s
+8\cS\de^{a}\de_{a}\s
\Big\}
\non\\
&&
+  \mbox{exact three-form}
~.
\label{sWvard2}
\eea

It is seen that each of the  closed three-forms constructed varies non-trivially under 
the super-Weyl transformations.
However, the crucial point is that, for any value of $\l$,  there exists 
 a linear combination of these three-forms which is  super-Weyl invariant
 modulo an exact three-form. One may check that 
the three-form
\bea
{\frak J}
={\bm \S}
-(1+\l)\frac{\pa{\bm \S} }{\pa\l}
+\frac{(3+2\l+\l^2)}{2}\frac{\pa^2{\bm \S}  }{\pa\l^2}
\label{cool-3-form}
\eea
possesses the following properties:
\begin{subequations}
\bea
\rd {\frak J}
&=&0~, \\
 \d_\s {\frak J}
 &=& \mbox{exact three-form}
\label{4.37b}
~, \\
\frac{\pa \frak J }{\pa\l} &=&0~.
\eea
\end{subequations}
To prove \eqref{4.37b} we can proceed in two steps.
First, we check that the super-Weyl 
variation of ${\frak  J}_{a\b\g}$ can be represented as
\bea
\d_\s{\frak  J}_{a\b\g}=
8(\g_a)_{\b\g}\de^\a(\cS\de_\a\s)
+\ve_{abc}(\g^b)_{\b\g}V^c~,
\eea
where $V^c$ is a vector that  can easily be computed using the relations 
\eqref{sWvard0}--\eqref{cool-3-form}.
Next, 
making use of 
eqs. \eqref{veV}--\eqref{deV} and ignoring exact terms,
 we observe that 
 $\d_\s {\frak J}$ is an exact three-form.

Let us write down the final expression for $\frak J$
\bea
{\frak J}&=&
\frac{1}{6}E^\g E^\b E^\a \Big\{
 \ve_{abc}{\bm\O}_\a{}^a{\bm\O}_\b{}^b{\bm\O}_\g{}^c
\Big\}
\non\\
&&
+\frac{1}{2}E^\g E^\b E^a\Big\{
16\ri(\g_a)_{\b\g}\cS^2
+2(1+\l)\ve_{acd}\cS{\bm\O}_\b{}^c {\bm\O}_\g{}^d
+\ve_{bcd}{\bm\O}_a{}^b  {\bm\O}_\b{}^c {\bm\O}_\g{}^d
\non\\
&&~~~~~~~~~~~~~~~~~
+\Big{[}\,
\frac{8}{3}(\de_{\b}\cS)\d_a^c
-\frac{8}{3}\ve_{ab}{}^{c}(\g^b)_{\b}{}^{\a}(\de_{\a}\cS)
+2(\g_a)_{\b\a}(\g^c)_{\d\r}C^{\a\d\r}
\Big{]}{\bm\O}_\g{}_c
\Big\}
\non\\
&&
+\frac{1}{2}E^\g E^b E^a\ve_{abc}
\Big\{
-2(1+\l)(\g^c)^{\g\d}C_{\b\g\d}\cS
+\frac{16(2-\l)}{3}(\g^c)_{\b}{}^\g(\de_{\g}\cS)\cS
\non\\
&&~~~~~~~~~
+\Big{[}\frac{\ri}{2}(\g^c)^{\a\b}(\g^e)^{\d\r}\de_{(\a}C_{\b\d\r)}
+\Big(\frac{2\ri}{3}(\de^2\cS)+4(1+\l)^2\cS^2\Big)\eta^{ce}
+2\ve^{cde}(\de_{d}\cS)
\Big{]}{\bm\O}_\g{}_e
\non\\
&&~~~~~~~~~
+\Big{[}
\ve^{cde}(\g_{e})_\g{}^{\a}(\g^f)^{\d\r}C_{\a\d\r}
-\frac{4}{3}(\de_{\g}\cS)\ve^{cdf}
+\frac{8}{3}\eta^{f[c}(\g^{d]})_{\g}{}^{\a}(\de_{\a}\cS)
\Big{]}{\bm\O}_{d}{}_f  
\non\\
&&~~~~~~~~~
-4(1+\l)\cS{\bm\O}_e{}^{[c}  {\bm\O}_\g{}^{e]}
-\frac{1}{2}\ve^{cde} \ve_{fgh}{\bm\O}_d{}^f  {\bm\O}_e{}^g  {\bm\O}_\g{}^h
\Big\}
\non\\
&&
+\frac{1}{6}E^cE^bE^a\ve_{abc}\Big\{
2\ri C^{\mu\r\t}C_{\mu\r\t}
-\frac{8\ri}{3}(\de^{\g}\cS)(\de_{\g}\cS)
-4\ri(1-\l)(\de^2\cS)\cS
\non\\
&&~~~~~~~~~
-8(3-3\l-3\l^2-\l^3)\cS^3
\non\\
&&~~~~~~~~~
+\Big{[}\frac{\ri}{2}(\g^d)^{\a\b}(\g^f)^{\g\d}\de_{(\a}C_{\b\g\d)}
+\Big(\frac{2\ri}{3}(\de^2\cS)+4(1+\l)^2\cS^2\Big)\eta^{df}
+2\ve^{def}(\de_{e}\cS)
\Big{]}{\bm\O}_d{}_f  
\non\\
&&~~~~~~~~~
+2(1+\l)\cS{\bm\O}_{d}{}^{[d}  {\bm\O}_{e}{}^{e]}  
-\frac{1}{6}\ve^{e_1e_2e_3}\ve_{d_1 d_2 d_3}
{\bm\O}_{e_1}{}^{d_1}  {\bm\O}_{e_2}{}^{d_2}  {\bm\O}_{e_3}{}^{d_3}  
\Big\}
~.
\eea

Since the three-form $\frak J$ is $\l$-independent, 
any convenient value of $\l$ may be used in order to 
compute $\frak J$. Setting $\l=0$ gives 
\bea
{\frak J} &=&
\frac{1}{6}E^\g E^\b E^\a \Big\{
 \ve_{abc}\O_\a{}^a\O_\b{}^b\O_\g{}^c
\Big\}
\non\\
&&
+\frac{1}{2}E^\g E^\b E^a\Big\{
16\ri(\g_a)_{\b\g}\cS^2
+2\ve_{acd}\cS\O_\b{}^c \O_\g{}^d
+\ve_{bcd}\O_a{}^b\O_\b{}^c \O_\g{}^d
\non\\
&&~~~~~~~~~~~~~~~~~
+\Big{[}\,
\frac{8}{3}\ve_{ab}{}^{c}(\g^b)_{\b\a}(\cD^{\a}\cS)
+\frac{8}{3}(\cD_{\b}\cS)\d_a^c
-2(\g_a)_\b{}^{\a}(\g^c)^{\d\r}C_{\a\d\r}
\Big{]}\O_\g{}_c
\Big\}
\non\\
&&
+\frac{1}{2}E^\g E^b E^a\ve_{abc}
\Big\{
-2(\g^c)^{\d\r}C_{\g\d\r}\cS
+\frac{32}{3}(\g^c)_{\g}{}^\d(\cD_{\d}\cS)\cS
\non\\
&&~~~~~~~~~~~~~~~
+\Big{[}\frac{\ri}{2}(\g^c)^{\a\b}(\g^e)^{\d\r}\cD_{(\a}C_{\b\d\r)}
+\Big(\frac{2\ri}{3}(\cD^2\cS)+4\cS^2\Big)\eta^{ce}
+2\ve^{cde}(\cD_{d}\cS)
\Big{]}\O_\g{}_e
\non\\
&&~~~~~~~~~~~~~~~
+\Big{[}
\ve^{cde}(\g_{e})_\g{}^{\a}(\g^f)^{\d\r}C_{\a\d\r}
-\frac{4}{3}(\cD_{\g}\cS)\ve^{cdf}
+\frac{8}{3}\eta^{f[c}(\g^{d]})_{\g}{}^{\a}(\cD_{\a}\cS)
\Big{]}\O_{df}
\non\\
&&~~~~~~~~~~~~~~~
-4\cS\O_e{}^{[c}\O_\g{}^{e]}
-\frac{1}{2}\ve^{cde} \ve_{fgh}\O_d{}^f\O_e{}^g\O_\g{}^h
\Big\}
\non\\
&&
+\frac{1}{6}E^cE^bE^a\ve_{abc}\Big\{
2\ri C^{\a\r\t}C_{\a\r\t}
-\frac{8\ri}{3}(\cD^{\g}\cS)(\cD_{\g}\cS)
-4\ri(\cD^2\cS)\cS
-24\cS^3
\non\\
&&~~~~~~~~~~~~~~~
+\Big{[}\frac{\ri}{2}(\g^d)^{\a\b}(\g^f)^{\g\d}\cD_{(\a}C_{\b\g\d)}
+\Big(\frac{2\ri}{3}(\cD^2\cS)+4\cS^2\Big)\eta^{df}
+2\ve^{def}(\cD_{e}\cS)
\Big{]}\O_{df}
\non\\
&&~~~~~~~~~~~~~~~
+2\cS\O_{d}{}^{[d}\O_{e}{}^{e]}
-\frac{1}{6}\ve^{e_1e_2e_3}\ve_{d_1 d_2 d_3}
\O_{e_1}{}^{d_1}\O_{e_2}{}^{d_2}\O_{e_3}{}^{d_3}
\Big\}
~.
\label{cool-result-1}
\eea

\subsection{The action principle for conformal supergravity} 

The three-form $\frak J$ is our main result. Associated with $\frak J$ is the action 
for $\cN=1$ conformal supergravity defined via the integration of $\frak J$ over spacetime $\cM_3$:
\bea
S_{\rm CSG} = \int_{\cM_3} {\frak J} = \int \rd^3 x \, ({}^* {\frak J}) ~, \qquad 
{}^*{\frak J} = \frac{1}{3!} \ve^{mnp}{\frak J}_{mnp} ~.
\label{S-CSG}
\eea
The action is automatically invariant under the local Lorentz and super-Weyl transformations, 
since the corresponding variations of 
the three-form $\frak J$ have been shown to be exact. 
It remains to show that $S_{\rm CSG} $ is invariant under general coordinate transformations 
of the curved superspace generated by the vector field  $\x^C E_C $ in 
\eqref{SUGRA-gauge-group1}. 
It suffices to repeat the four-dimensional proof due to Hasler \cite{Hasler} (see also \cite{GGKS})
\bea
\d_\x {\frak J} = \cL_\x {\frak J}= \imath_\x \,\rd \, {\frak J} + \rd \,\imath_\x \,{\frak J} 
=\rd \,\imath_\x \,{\frak J} ~,
\eea
where $ \imath_\x $ denotes the interior product and $\cL_\x$ the Lie derivative. 
Since the variation $\d_\x {\frak J} $ is an exact three-form, 
the action $S_{\rm CSG} $ is indeed invariant under superdiffeomorphisms provided 
$\cM_3$ has no boundary.

\section{Component action}
\setcounter{equation}{0}

In this section we reduce 
the action \eqref{S-CSG}
to the component fields, choosing a special Wess-Zumino gauge,
and demonstrate that 
it coincides with 
the well-known action for $\cN=1$ conformal supergravity \cite{vN}.
To start with, we elaborate on the component reduction.

\subsection{Components reduction}

Given a superfield $U(z)$ we define its bar-projection $U|$ to be 
the $\q$-independent term in the expansion of $U(x,\q)$ in powers of $\q$'s,
\bea
U|:=U(x,\q)|_{\q=0}~.
\eea
In a similar way we  define the bar-projection of the covariant derivatives: 
\bea
\cD_A|:=E_A{}^M |\pa_M+\hf\O_A{}^{bc}|\cM_{bc}~.
\eea

The supergravity gauge freedom may be used to algebraically 
gauge away a number of component fields contained in $\cD_A$ except those
which constitute 
the Weyl multiplet of conformal supergravity.
The supergravity gauge group is spanned by the general coordinate, local Lorentz and 
super-Weyl transformations. 

The freedom to perform  general coordinate and local 
SL(2,$\mathbb{R}$) transformations  
can be used to choose a Wess-Zumino gauge of the form 
\bsubeq\label{WZ}
\bea
&&\cD_\a|=\d_\a{}^\m\frac{\pa}{\pa\q^\m}\quad \Longleftrightarrow \quad
E_\a{}^{\mu}|=\d_\a{}^\mu~,~~
\O_\a{}^{bc}|=0
\label{WZ-1}
~,
\\
&&\cD_a|=\bD_a +{\Psi}_a{}^\g(x)\cD_\g|
~,
\eea
\esubeq
where $\bD_a$ denotes a space-time covariant derivative
\bea
\bD_a=e_a+\o_a~,~~~~~~
e_a=e_a{}^m(x)\pa_m~,~~
\o_a=\hf\o_a{}^{bc}(x)\cM_{bc} ~.
\eea
Here the component inverse vielbein $e_a{}^m(x)$ and the 
component vielbein $e_m{}^a(x)$ are defined as
\bea
e_a{}^m:=E_a{}^m|~,\qquad
e_m{}^a:=E_m{}^a|~.
\eea
They are related to each other in the standard way 
\bea
e_a{}^me_m{}^b=\d_a^b~,\qquad
e_m{}^ae_a{}^n=\d_m^n~.
\eea
The component Lorentz connection is defined as
\bea
\o_{a}{}^{bc}:=\O_a{}^{bc}|~.
\eea
Finally, the gravitino is defined by the rule:
\bea
\Psi_a{}^\g:=E_a{}^\g|
~,\qquad 
e_m{}^a\Psi_a{}^\g := -E_m{}^\g|
~.
\label{gravitino}
\eea

The space-time covariant derivatives $\bD_a$ obey the commutation relations
\bea
[\bD_a,\bD_b]&=&\cT_{ab}{}^c \bD_c
+\hf\cR_{ab}{}^{cd}\cM_{cd}
~,
\eea
with $\cT_{ab}{}^c$ the torsion  and $\cR_{ab}{}^{cd}$ the curvature.
Their explicit  expressions are
\bsubeq
\bea
\cT_{ab}{}^c&=&\cC_{ab}{}^c+2\o_{[a}{}_{b]}{}^c~,
\\
\cR_{ab}{}^{cd}&=&
2e_{[a}\o_{b]}{}^{cd}
+2\o_{[a}{}_{b]}{}^f\o_f{}^{cd}
+2\o_{[a}{}^{cf}\o_{b]}{}_f{}^d
-\cT_{ab}{}^f\o_f{}^{cd}
~,
\eea
\esubeq
where $\cC_{ab}{}^c$ stands for the anholonomy coefficients, 
\bea
[e_a , e_b] = \cC_{ab}{}^c e_c~, \qquad  
\cC_{ab}{}^c=2(e_{[a}e_{b]}{}^n)e_n{}^c~.
\eea
The connection is uniquely determined as a function of the vielbein and torsion, $\o_a{}^{cd}=\o_a{}^{cd}(e,\cT)$. Its explicit form is 
\bea
\o_{abc}=\hf\Big{[}(\cT_{abc}-\cC_{abc})
-(\cT_{bca}-\cC_{bca})+(\cT_{cab}-\cC_{cab})\Big{]}~.
\eea

So far we have partially fixed the general coordinate and local Lorentz
symmetries. We still have the complete super-Weyl gauge freedom at our disposal.
Let us recall 
the super-Weyl transformation of the torsion  superfield $\cS$, eq. 
\eqref{sWS}.  
It follows from \eqref{sWS} that we are in a position to choose the gauge
\bea
\cS| =0
~.
\label{WZ-2}
\eea
The conditions \eqref{WZ} and \eqref{WZ-2} constitute the complete Wess-Zumino guage. 
In this gauge,  
the only independent component fields are the vielbein and the gravitino, 
and they comprise  the Weyl multiplet of 
$\cN=1$ conformal supergravity.
The gauge condition \eqref{WZ} does not fix completely the general coordinate 
and local Lorentz symmetries of the curved superspace. The residual gauge transformations, 
which preserve the condition \eqref{WZ},  are spanned by 
(i) the general coordinate transformations in space-time; (ii) the 
local Lorentz transformations in space-time; and (iii) the $Q$-supersymmetry transformations. 
The gauge condition \eqref{WZ-2} only partially fixes the super-Weyl gauge freedom. 
The residual super-Weyl transformations, which preserve  \eqref{WZ-2}, 
are (iv) the space-time Weyl transformations; and (v) the $S$-supersymmetry transformations.

To complete the component reduction, we express
the gravitino field strength and the  component torsion and curvature 
in terms of the superfield torsion:
\bsubeq
\bea
\cT_{ab}{}^c&=&
-2\ri(\g^c)_{\g\d}{\Psi}_a{}^\g{\Psi}_b{}^\d
~,
\label{components-1}
\\
\ve_{c}{}^{ab}(\bD_{[a}{\Psi}_{b]}{}^\r)
&=&
\Big{[}
-\frac{\ri}{2} (\g_c)_{\a\b}C^{\r\a\b}
+\frac{2\ri}{3}(\g_c)^{\r\a}(\cD_{\a}\cS)
+\hf\ve_{c}{}^{ab}\cT_{ab}{}^d
{\Psi}_d{}^\r 
\Big{]}\Big|
~,~~~~~~~~~
\label{components-2}
\\
\ve^{abc}\cR_{bc}{}^{d}&=&
\Big{[}\,
\ri(\g^a)^{\a\b}(\g^d)^{\g\d}(\cD_{(\a}C_{\b\g\d)})
+\frac{4\ri}{3}(\cD^2\cS)\eta^{cd}
\non\\
&&~
+\ve^{abc}{\Psi}_{b}{}^\g\Big(
2(\g_c)_\g{}^\d(\g^d)^{\a\b}C_{\d\a\b}
+\frac{4}{3}\big(\d_\g^\d\d_c^d
+2\ve_{ce}{}^d(\g^e)_\g{}^\d\big)\cD_\d\cS
\Big)
\Big{]}\Big|
~.~~~~~~~~
\label{components-3}
\eea
\esubeq
Here we have denoted $\cR_{ab}{}^{d}=\hf\ve^d{}_{ef}\cR_{ab}{}^{ef}$.
Since the torsion is a quadratic polynomial of  the gravitino, eq. \eqref{components-1}, 
the Lorentz connection
is determined in terms of the vielbein and gravitino, 
$\o_a{}^{cd}=\o_a{}^{cd}(e,\Psi)$.

\subsection{The action for conformal supergravity in Wess-Zumino gauge}

Our superspace action for 
$\cN=1$ conformal supergravity, eq. \eqref{S-CSG},
is constructed in terms of 
the closed three-form $\frak J$ given by \eqref{cool-result-1}.
We now express it in terms of the component fields. 
The action can equivalently be rewritten in the following form
\bea
S_{\rm CSG}=\frac{1}{6}\int\rd^3x\,e\,
\ve^{abc}
\Big{[}
{\frak J}_{abc}
-3\Psi_a{}^\g\, {\frak J}_{bc\g}
-3\Psi_a{}^\b \Psi_b{}^\g\, {\frak J}_{c\b\g}
+\Psi_a{}^\a \Psi_b{}^\b \Psi_{c}{}^\g\,{\frak J}_{\a\b\g}
\Big{]}\Big|
~,
\eea
where 
$e={\rm det}{(e_m{}^a)}$ and the gravitino $\Psi_a{}^\g$ is defined according to 
\eqref{gravitino}.  
To compute the integrand, we have to make use of the explicit expressions for 
the components of the 
three-form $\frak J$, eq. \eqref{cool-result-1}.
In the Wess-Zumino gauge defined by eqs. \eqref{WZ-1} and  \eqref{WZ-2},
the result is 
\bea
S_{\rm CSG} &=&
\int\rd^3x\,e\,
\Big{\{}
-2\ri C^{\a\b\g}C_{\a\b\g}
+\frac{8\ri}{3}(\cD^{\a}\cS)\cD_{\a}\cS
+\frac{1}{6}\ve^{a_1a_2a_3}\ve_{b_1 b_2 b_3}\o_{a_1}{}^{b_1} \o_{a_2}{}^{b_2} \o_{a_3}{}^{b_3} 
\non\\
&&~~~~~~
+\Psi_a{}^\a\Big{[}
\ve^{abd}(\g_{d})_\a{}^{\b}(\g^c)^{\g\d}C_{\b\g\d}
+\frac{2}{3}(\cD_{\a}\cS)\ve^{abc}
+\frac{8}{3}\eta^{c[a}(\g^{b]})_{\a}{}^{\b}(\cD_{\b}\cS)
\Big{]}
\o_{bc} 
\non\\
&&~~~~~~
+\Big{[}
-\frac{\ri}{2}(\g^a)^{\a\b}(\g^b)^{\g\d}\cD_{(\a}C_{\b\g\d)}
-\frac{2\ri}{3}(\cD^2\cS)\eta^{ab}
\Big{]}\o_{ab} 
\Big{\}}\Big|
~.
\label{action-WZ}
\eea
It only remains to make use of 
the relations \eqref{components-2} and \eqref{components-3}
to arrive at the final expression for the action:
\bea
S_{\rm CSG}  &=&
\frac{1}{4}\int\rd^3x\,e\,  \ve^{abc}
\Big{\{}
  \o_{a}{}^{fg}\cR_{bc}{}_{fg}
+\frac{2}{3}\o_{a}{}_{f}{}^{g} \o_{b}{}_{g}{}^{h}\o_{c}{}_h{}^{f}
\non\\
&&
-8\ri
\Big(\bD_b{\Psi}_c{}^\a-\hf\cT_{bc}{}^{g}{\Psi}_g{}^\a\Big)
(\g_d)_\a{}^\b(\g_a)_\b{}^\g\ve^{def}\Big(\bD_e{\Psi}_f{}_\g-\hf\cT_{ef}{}^h{\Psi}_h{}_\g\Big)
\Big{\}}
~.~~~~~~~~~
\eea
Up to an overall factor of $1/4$, 
this is the well-known action for $\cN=1$ conformal supergravity 
\cite{vN}.

\section{Outlook}
\setcounter{equation}{0}

In this paper we have presented the new superfield method to construct the action for three-dimensional
$\cN=1$ conformal supergravity, and thus for $\cN=1$ topologically massive supergravity. 
The power of this method is that it may naturally be generalized to the case of $\cN$-extended conformal supergravity. Here we only sketch such a generalization, leaving details for a future publication. 

Let $\cD_A =(\cD_a, \cD_\a^I)$ be the superspace covariant derivatives, with $I=1,\dots,  \cN$, 
which describe the off-shell $\cN$-extended Weyl supermultiplet \cite{HIPT,KLT-M11}. 
Following the conventions of \cite{KLT-M11}, we consider a
two-parameter deformation of the vector covariant derivative
\bea
\cD_{\a\b} ~\to ~ \nabla_{\a\b} = \cD_{\a\b} +\l \cS \cM_{\a\b} + \r C_{\a\b}{}^{KL} \cN_{KL}~,
\eea
where $\l$ and $\r$ are real parameters, and $\cS$ and $  C_{\a\b}{}^{KL} $ are certain dimension-one
torsion tensors.  
The deformed covariant derivatives $\nabla_A =(\nabla_a, \nabla_\a^I)
:= (\nabla_a, \cD_\a^I)$ 
obey the algebra
\bea
{[}\nabla_{{A}},\nabla_{{B}}\}&=&{\bm T}_{{A}{B}}{}^{{C}}\nabla_{{C}}
+\hf {\bm R}_{{A}{B}}{}^{cd}\cM_{cd}
+\hf {\bm R}_{AB}{}^{KL}\cN_{KL}~,
\eea
with  ${\bm T}_{AB}{}^C$ the torsion,  ${\bm R}_{AB}{}^{cd}$ the Lorentz curvature and
 ${\bm R}_{AB}{}^{KL}$ the SO($\cN$) curvature.
As a next stage, we have  to consider the equation
\bea
\rd {\bm \S}= \hf {\bm R}^{ab} {\bm R}_{ab} + \frac{\k}{2}  {\bm R}^{IJ} {\bm R}_{IJ} ~,
\label{6.3}
\eea
with $\k$  a real parameter, and look for two solutions ${\bm \S}_{\rm T}$ and ${\bm \S}_{\rm CS}$.
Here ${\bm \S}_{\rm T}$ is a three-form constructed in terms of the torsion and curvature 
tensors and their covariant derivatives, while   ${\bm \S}_{\rm CS}$ is a standard Chern-Simons three-form. 
Now, the three-form ${\bm \S} := {\bm \S}_{\rm T} - {\bm \S}_{\rm CS}$ has the following properties
(i) ${\bm \S} $  is closed;  and (ii) ${\bm \S} $ is 
a polynomial in two variables $\l$ and $\r$. By differentiating $\bm \S$ with respect to $\l$ and $\r$, 
we will generate
 a number of closed three-forms. Finally, we have to look for a linear combinations 
$\frak J$ of these closed three-forms, which is super-Weyl  invariant  modulo exact contributions. 
The parameter $\k$ is expected to be fixed by this requirement. 
It is also expected that ${\frak J}$ is independent 
of $\lambda$ and $\rho$, due to its uniqueness. 
The closed three-form $\frak J$ generates the action for $\cN$-extended conformal supergravity. 
\\

\noindent
{\bf Acknowledgements:}\\
We thank U.~Lindstr\"om, M.~Ro\v{c}ek and I.~Sachs for discussions that rekindled our interest in 
the problem solved in this paper. We are also grateful to Joseph Novak for reading the manuscript. 
The work of SMK was supported in part by the Australian Research Council
under Grant No. DP1096372.
The work of GT-M was supported by the Australian Research Council's Discovery Early Career 
Award (DECRA), project No. DE120101498.

\appendix

\section{Supersymmetric action}
\setcounter{equation}{0}

For completeness of our presentation, in this appendix we review the structure
of the locally supersymmetric action  \cite{Becker:2003wb}
associated with the closed three-form \eqref{2.21}.
This action is defined, in complete  analogy with \eqref{S-CSG},
by integrating the three-form $\X$ over the space-time,
 \bea
S (\cL)  = \int_{\cM_3} \X = \int \rd^3 x \, ({}^* \X) ~, \qquad 
{}^*\X = \frac{1}{3!} \ve^{mnp}\X_{mnp} ~,
\eea
in accordance with  the superform approach to  construct supersymmetric invariants 
\cite{Hasler,Ectoplasm,GGKS}.
In the Wess-Zumino gauge \eqref{WZ-1}, the action can readily be  
brought to the form
\bsubeq
\bea
S(\cL)&=&
\frac{1}{6}\int\rd^3x\,e\,
\ve^{abc}
\Big{[}\,
\X_{abc}
-3\Psi_a{}^\g\, \X_{bc\g}
-3\Psi_a{}^\b \Psi_b{}^\g\, \X_{c\b\g}
+\Psi_a{}^\a \Psi_b{}^\b \Psi_{c}{}^\g\,\X_{\a\b\g}
\Big{]}
~,~~~~~~
\label{A.2a}
\eea
or equivalently
\bea
S(\cL)&=&\int\rd^3x\,e\,
\Big{[}
\dfrac{\ri}{4}\cD^2+2\cS
-\hf(\g^a)^{\g\d}\Psi_a{}_\g \cD_{\d}
-\dfrac{\ri}{2}\ve^{abc}\Psi_a{}^\b \Psi_b{}^\g(\g_c)_{\b\g}
\Big{]}
\cL
~.
\label{L-comp-ac}
\eea
\esubeq
The locally supersymmetric action \eqref{A.2a} was derived in  \cite{Becker:2003wb}.
Our consideration in section \ref{section2}
shows that this action is super-Weyl invariant. 
The component action \eqref{L-comp-ac} 
was first derived in \cite{LR-brane}.

\section{Exact three-forms}
\setcounter{equation}{0}

Given a two-form ${\bm F}_2=\hf E^BE^A{\bm F}_{AB}$, its exterior derivative is
\bea
\rd {\bm F}_2&=&
\hf E^CE^BE^A\Big\{\de_{A}{\bm F}_{BC}-{\bm T}_{AB}{}^D{\bm F}_{DC}\Big\}
~.
\eea
Making use of the explicit expression for the torsion associated with the covariant derivatives
$\nabla_A$, we obtain
\bea
\rd {\bm F}_2&=&
\frac{1}{6}E^\g E^\b E^\a\Big\{3\de_{\a}{\bm F}_{\b\g}-6\ri(\g^d)_{\a\b}{\bm F}_{d\g}\Big\}
\non\\
&&
+\hf E^\g E^\b E^a\Big\{\de_{a}{\bm F}_{\b\g}
-2\de_{\b}{\bm F}_{a\g}
-2(1-\l)(\g_a)_{(\b}{}^\d {\bm F}_{\g)\d}\cS
-2\ri(\g^c)_{\b\g}{\bm F}_{ca}
\Big\}
\non\\
&&
+\hf E^\g E^bE^a\ve_{abc}\Big\{
-\ve^{cde}\de_{d}{\bm F}_{e\g}
-\hf\ve^{cde}\de_{\g}{\bm F}_{de}
-4\l {\bm F}^c{}_{\g}\cS
+(1-\l)\ve^{cab}(\g_a)_{\g}{}^\d {\bm F}_{b\d}\cS
\non\\
&&~~~~~~~~~~~~~~~~~~~~~~
+\Big{[}\frac{2\ri}{3}(\g^c)^{\a\d}(\de_{\a}\cS)
-\frac{\ri}{2}(\g^c)_{\a\b}C^{\a\b\d}\Big{]}
{\bm F}_{\g\d}
\Big\}
\non\\
&&
+\frac{1}{6} E^cE^bE^a\ve_{abc}\Big\{
-\hf\ve^{def}\de_{d}{\bm F}_{ef}
+\Big{[}\frac{\ri}{2}(\g^d)_{\a\b}C^{\a\b\d}
-\frac{2\ri}{3}(\g^d)^{\a\d}(\de_{\a}\cS)\Big{]}
{\bm F}_{d\d}
\Big\}
~.
\label{dF}
\eea

The relation  (\ref{dF}) provides us with a rule for ``integration by parts'' within the superform 
approach to constructing supersymmetric actions in three dimensions. Such an action 
is associated with a closed three-form, which plays the role of the Lagrangian. 
The action does not change if the Lagrangian is shifted by an exact three-form
provided we are allowed to ignore boundary terms
(in other words, all Lagrangians in the same cohomology class define the same action).  
In carrying out the explicit calculations for this paper, we have found extremely useful 
the following two special cases of the rule for ``integration by parts''  (\ref{dF}).

Firstly, given an antisymmetric tensor ${\bm F}_{ab}$, which is  equivalent to the vector 
${\bm F}^a =\hf \ve^{abc}{\bm F}_{bc}$,
it follows from (\ref{dF}) that
\bea
\hf E^\g E^\b E^a\Big\{
\ve_{abc}(\g^b)_{\b\g}{\bm F}^c
\Big\}
&=&
\hf E^\g E^bE^a\ve_{abc}\Big\{
\frac{\ri}{2}\de_{\g}{\bm F}^c
\Big\}
+\frac{1}{6} E^cE^bE^a\ve_{abc}\Big\{
\frac{\ri}{2}\de_d{\bm F}^d
\Big\}
\non\\
&&
+\mbox{exact three-form}
~.
\label{veV}
\eea

Secondly, 
suppose that 
${\bm F}_{a\b}=(\g_a)_{\b\g}{\bm V}^\g$,  
for some spinor ${\bm V}^\g$.
Then it follows from 
  \eqref{dF} and \eqref{veV} that 
\bea
\hf E^\g E^\b E^a\Big\{
(\g_a)_{\b\g}\de^{\a}{\bm V}_\a
\Big\}
&=&
\hf E^\g E^bE^a\ve_{abc}\Big\{
-\frac{3}{2}(1+\l) (\g^c)_{\g}{}^{\a}{\bm V}_\a\cS
-\frac{\ri}{4}(\g^c)_\g{}^{\a}\de^2{\bm V}_\a
\non\\
&&~~~~~~~~~~~~~~~~~~~~
+\hf\ve^{cde}(\g_{d})_{\g}{}^{\a}\de_{e}{\bm V}_\a
-\frac{1}{2}\de^c{\bm V}_\g
\Big\}
\non\\
&&
+\frac{1}{6} E^cE^bE^a\ve_{abc}\Big\{
2\ri(\de^{\a}\cS){\bm V}_\a
-\frac{\ri}{2}(\g^d)^{\r\t}\de_{d}\de_{\r}{\bm V}_\t
\Big\}
\non\\
&&
+\mbox{exact three-form}
~.
\label{deV}
\eea


\begin{footnotesize}

\end{footnotesize}


\begin{thebibliography}{66}

\bibitem{DK} 
S.~Deser and J.~H.~Kay,
``Topologically massive supergravity,''
Phys.\ Lett.\ B {\bf 120}, 97 (1983).

\bibitem{Deser}
  S.~Deser,
  ``Cosmological topological supergravity,''
 in {\it Quantum Theory Of Gravity}, S. M. Christensen (Ed.), 
 Adam Hilger, Bristol, 1984, pp. 374-381. 



\bibitem{vN}
P.~van Nieuwenhuizen,
``D = 3 conformal supergravity and Chern-Simons terms,''
Phys.\ Rev.\  D {\bf 32}, 872 (1985).

\bibitem{RvN}
M.~Ro\v{c}ek and P.~van Nieuwenhuizen,
``${\rm N}   \geq 2$ supersymmetric Chern-Simons terms as D = 3 extended conformal
supergravity,''
Class.\ Quant.\ Grav.\  {\bf 3}, 43 (1986).

\bibitem{LR89}
  U.~Lindstr\"om and M.~Ro\v{c}ek,
  ``Superconformal gravity in three dimensions as a gauge theory,''
  Phys.\ Rev.\ Lett.\  {\bf 62}, 2905 (1989).
  

  
\bibitem{GGRS}
 S.~J.~Gates, Jr., M.~T.~Grisaru, M.~Ro\v{c}ek and W.~Siegel,
{\it Superspace, or One Thousand and One Lessons in Supersymmetry},
Front.\ Phys.\  {\bf 58}, 1 (1983) [arXiv:hep-th/0108200].

\bibitem{ZP88}
  B.~M.~Zupnik and D.~G.~Pak,
   ``Superfield formulation of the simplest three-dimensional gauge theories and
  conformal supergravities,''
  Theor.\ Math.\ Phys.\  {\bf 77} (1988) 1070
  [Teor.\ Mat.\ Fiz.\  {\bf 77} (1988) 97].
  
\bibitem{ZP89} 
B.~M.~Zupnik and D.~G.~Pak,
``Differential and integral forms in supergauge theories and supergravity,''
Class.\ Quant.\ Grav.\  {\bf 6}, 723 (1989).
  

\bibitem{HIPT}
  P.~S.~Howe, J.~M.~Izquierdo, G.~Papadopoulos and P.~K.~Townsend,
  ``New supergravities with central charges and Killing spinors in 2+1 dimensions,''
  Nucl.\ Phys.\  B {\bf 467}, 183 (1996)
  [arXiv:hep-th/9505032].

\bibitem{KLT-M11}
S.~M.~Kuzenko, U.~Lindstr\"om and G.~Tartaglino-Mazzucchelli,
  ``Off-shell supergravity-matter couplings in three dimensions,''
  JHEP {\bf 1103}, 120 (2011)
  [arXiv:1101.4013 [hep-th]].
  
\bibitem{BKN} 
D.~Butter, S.~M.~Kuzenko and J.~Novak,
``The linear multiplet and ectoplasm,''
  JHEP {\bf 1209}, 131 (2012)
  [arXiv:1205.6981 [hep-th]].

\bibitem{Hasler} 
M.~F.~Hasler,
``The three-form multiplet in N=2 superspace,''
Eur.\ Phys.\ J.\ C {\bf 1}, 729 (1998) [hep-th/9606076].

\bibitem{Ectoplasm} 
S.~J.~Gates, Jr., ``Ectoplasm has no topology: The prelude,''
in {\it Supersymmetries and Quantum Symmetries},
 J. Wess and E. A. Ivanov (Eds.), Springer, Berlin, 1999, p. 46, arXiv:hep-th/9709104;
``Ectoplasm has no topology,''
 Nucl.\ Phys.\  B {\bf 541}, 615 (1999)
 [arXiv:hep-th/9809056].

\bibitem{GGKS}
S.~J.~Gates, Jr., M.~T.~Grisaru, M.~E.~Knutt-Wehlau and W.~Siegel,
``Component actions from curved superspace: Normal coordinates and
ectoplasm,'' Phys.\ Lett.\  B {\bf 421}, 203 (1998)
[hep-th/9711151].

\bibitem{HT}
  P.~S.~Howe and R.~W.~Tucker,
  ``A locally supersymmetric and reparametrization invariant action for a
  spinning membrane,''  J.\ Phys.\ A  {\bf 10}, L155 (1977);
  ``Local supersymmetry in (2+1) dimensions. 1. Supergravity and differential forms,''
  J.\ Math.\ Phys.\  {\bf 19}, 869 (1978);
  ``Local supersymmetry in (2+1) dimensions. 2. An action for a spinning membrane,''
  J.\ Math.\ Phys.\  {\bf 19}, 981 (1978).




  \bibitem{BG}
  M.~Brown and S.~J.~Gates Jr.,
``Superspace Bianchi identities and the supercovariant derivative,''
  Annals Phys.\  {\bf 122}, 443 (1979).


\bibitem{Uematsu}
  T.~Uematsu,
  ``Structure of N=1 conformal and Poincare supergravity in (1+1)-dimensions
  and (2+1)-dimensions,''
  Z.\ Phys.\  C {\bf 29}, 143 (1985);
``Constraints and actions in two-dimensional and three-dimensional N=1
conformal supergravity,''
Z.\ Phys.\  C {\bf 32}, 33 (1986).

\bibitem{LR-brane}
  U.~Lindstr\"om and M.~Ro\v{c}ek,
  ``A super-Weyl-invariant spinning membrane,''
  Phys.\ Lett.\  B {\bf 218}, 207 (1989).

\bibitem{K12} 
  S.~M.~Kuzenko,
  ``Prepotentials for N=2 conformal supergravity in three dimensions,''
 JHEP {\bf 1212} (2012) 021 [arXiv:1209.3894 [hep-th]].

\bibitem{Becker:2003wb}
 M.~Becker, D.~Constantin, S.~J.~Gates Jr., W.~D.~Linch III, W.~Merrell and J.~Phillips,
 ``M-theory on Spin(7) manifolds, fluxes and 3D, N = 1 supergravity,''
  Nucl.\ Phys.\  B {\bf 683}, 67 (2004)
  [arXiv:hep-th/0312040].

\bibitem{Dragon}
N.~Dragon,
``Torsion and curvature in extended supergravity,''
Z.\ Phys.\  C {\bf 2}, 29 (1979).

\bibitem{Novak} 
  J.~Novak,
  ``Superform formulation for vector-tensor multiplets in conformal supergravity,''
  JHEP {\bf 1209}, 060 (2012)
  [arXiv:1205.6881 [hep-th]].

\end{thebibliography}
\end{document}